\documentclass[reprint,author-numerical,nofootinbib]{revtex4-2}
\usepackage[utf8]{inputenc}
\usepackage[english]{babel}
\usepackage{graphicx}% Include figure files
\usepackage{dcolumn}% Align table columns on decimal point
\usepackage{bm}% bold math
\usepackage{amsmath}
\usepackage{amsfonts}
\usepackage{amssymb}

\newtheorem{proposition}{Proposition}

\def\dif{{\rm d}}

\def\Remarks{\ \\{\bf Remarks }}
\newcommand{\be}{\begin{equation}}
\newcommand{\ee}{\end{equation}}

\begin{document}

\preprint{AIP/123-QED}

\title[Thermodynamic isotropic universes]{Thermodynamics of the universes admitting isotropic radiation}
%
%\title[Thermodynamic Stephani Universes]{Thermodynamics of the Stephani Universes}

\author{Salvador Mengual}
\email{salvador.mengual@uv.es.}
\affiliation{ Departament d'Astronomia i Astrof\'{\i}sica,
Universitat de Val\`encia, E-46100 Burjassot, Val\`encia, Spain.}
\author{Joan Josep Ferrando}
\altaffiliation[Also at ]{Observatori Astron\`omic, Universitat
de Val\`encia,  E-46980 Paterna, Val\`encia, Spain.}
\affiliation{ Departament d'Astronomia
i Astrof\'{\i}sica, Universitat
de Val\`encia, E-46100 Burjassot, Val\`encia, Spain.}%\\This line break forced with \textbackslash\textbackslash
\author{Juan Antonio S\'aez}
\affiliation{ Departament de Matem\`atiques per a l'Economia i
l'Empresa, Universitat de Val\`encia, E-46022 Val\`encia, Spain.}

\date{\today}

\begin{abstract}
The thermodynamic interpretation of the Stephani Universes is studied in detail. The general expression of the speed of sound and of the thermodynamic schemes associated with a thermodynamic solution is obtained. The constraints imposed on the solutions by considering some significant physical properties are analyzed. We focus on the models where the cosmological observer measures isotropic radiation. We consider some examples, and a solution that models an ultrarelativistic gas is analyzed in detail. 
\end{abstract} 

\maketitle

%%%%%%%%%%%%%%%%%%%%%%%%%%%%%%%%%%%%%%%
\section{\label{sec-intro}Introduction}
%%%%%%%%%%%%%%%%%%%%%%%%%%%%%%%%%%%%%%%

The perturbation theory of the Friedmann-Lema\^itre-Robertson-Walker (FLRW) models seems to provide a good explanation of the observed degree of inhomogeneity in the Universe. However, the structure and evolution of galaxies, clusters and voids require an analysis outside of the perturbative regime. They are often 
modeled by Newtonian N-body computations. But the nonlinear effects of the Einstein field equations could be critical in the structure formation, and a large number of studies have
been devoted to providing exact inhomogeneous models for studying the formation of structures and for analyzing the effect of the nonlinear inhomogeneities on the cosmic microwave background radiation (see \cite{Krasinski,Krasinski-Plebanski,Ellis-Maartens-MacCallum} and references therein).

A broad review of the inhomogeneous cosmological solutions, which contain the FLRW models as a limit, can be found in Krasi\'nski's book \cite{Krasinski}. Among others, two families of metrics are largely analyzed: the Szekeres-Szafron solutions \cite{Szekeres,Szafron} and the Stephani-Barnes metrics \cite{st,ba}. Krasi\'nski also remarks on the need to analyze if these inhomogeneous universes model realistic physical fluids.

Since then, considerable progress has been made in this direction. It is known that a thermodynamic Szekeres-Szafron solution of class I admits, necessarily, a three-dimensional group of isometries on two-dimensional orbits \cite{KQS} (see also the recent paper \cite{fs-SS}). These metrics are the Lema\^itre-Tolman models (with pressure) and their plane and hyperbolic counterparts. Recently \cite{MF-LT}, we have analyzed the thermodynamics of the subclass admitting a flat synchronization, and we have studied in depth the ideal gas models. 

On the other hand, thermodynamic Szekeres-Szafron solutions of class II without symmetries exist \cite{KQS}. The analysis of their thermodynamic properties and the study of the ideal gas models require distinguishing the singular and regular models \cite{CFS-PSS, CFS-RSS}. When the spacetime admits spherical, plane or hyperbolic symmetry, the solution is a T-model, whose thermodynamic properties have been analyzed recently \cite{FM-Tmodels, FM-Tmodels2}.
 
The conformally flat solutions of the Barnes-Stephani metrics are the {\em Stephani universes}, which were obtained by Stephani \cite{st}, and recovered later by Barnes \cite{ba} as the conformally flat class of irrotational and shear-free perfect fluid spacetimes with nonzero expansion. They can also be characterized as the spacetimes verifying a weak cosmological principle without any hypothesis on the energy tensor \cite{bc-0, bc}.

Bona and Coll \cite{bc} also showed that the necessary and sufficient condition for a Stephani universe to represent the evolution of a fluid in local thermal equilibrium is to admit a three-dimensional isometry group on two-dimensional orbits. This result was later recovered in \cite{KQS}, and spherically symmetric Stephani universes that may be interpreted either as a classical monoatomic ideal gas or as a matter-radiation mixture were considered in \cite{Sussman}.

In \cite{CF-Stephani} we studied the Stephani universes that can be interpreted as a generic ideal gas in local thermal equilibrium, and more recently \cite{CFS-CC} we have analyzed in depth the conditions for physical reality of the ideal models that approach a relativistic Synge gas at low or high temperatures. Despite these results, there are a lot of questions to deal with concerning the physical meaning of the Stephani universes. In order to understand these open problems that we analyze in this paper, we summarize below some notions on the necessary macroscopic conditions for physical reality and on the spacetimes admitting isotropic radiation.

%%%%%%%%%%%

\subsection{Macroscopic conditions for physical reality}
\label{subsec-necessary-conditions}

The evolution of a relativistic perfect fluid is expressed by an energy tensor in the form $T = (\rho+ p) u \otimes u + p \, g$, and fulfilling the divergence-free condition $\nabla \cdot T=0$. This constraint consists of a first-order differential system of four equations on five {\em hydrodynamic quantities}  ({\em unit velocity} $u$, {\em energy density} $\rho$, and {\em pressure} $p$).
%
%\be \label{ceq}
%\hspace{-0mm} {\rm C} :  \quad   \dif p  + u(p) u + (\rho + p) a = 0 \, ,  \quad 
%u(\rho) + (\rho+ p) \theta = 0 \, ,
%\ee
%
%where $a$ and $\theta$ are, respectively, the acceleration and the
%expansion of $u$, and where $u(q)$ denotes the directional
%derivative, with respect to $u$, of a quantity $q$, $u(q)
%= u^{\alpha} \partial_{\alpha} q$. 

%We are interested in perfect energy tensors $T$ that model realistic fluids when the thermodynamic perfect fluid approximation is suitable, that is, when the transport coefficients vanish (or are negligible) \cite{Eckart, Rezzolla}. Next, we summarize the complementary general macroscopic requirements that must be imposed on $T$ to represent the energetic evolution of a physically realistic perfect fluid (see the recent paper \cite{CFS-RSS} for more details).

{\em Energy conditions} are necessary algebraic conditions for physical reality and, in the perfect fluid case, they state \cite{Plebanski}: \\[-4.0mm]
\begin{equation} \label{e-c}
\hspace{-20mm} {\rm E} : \qquad   \qquad \qquad  -\rho < p \leq \rho  \, .
\end{equation}

If the energy tensor describes the (nonisoenergetic, $u(\rho) = u^\alpha \partial_\alpha \rho \not= 0$) evolution of a thermodynamic perfect fluid in {\em local thermal equilibrium}, then the hydrodynamic quantities $\{u, \rho, p\}$ must fulfill the {\em hydrodynamic sonic condition} \cite{Coll-Ferrando-termo, CFS-LTE}: \\[-5mm]
\begin{equation} \label{lte-chi}
\hspace{0mm} {\rm S} :  \qquad \quad     \dif \chi \wedge \dif p \wedge \dif \rho = 0 \, , \qquad \chi \equiv \frac{u(p)}{u(\rho)}   \, .
\end{equation}
When this condition holds, the {\em indicatrix of the local thermal equilibrium} $\chi$ is a function of state, $\chi = \chi(\rho,p)$, which physically represents the square of the {\em speed of sound} in the fluid, $\chi (\rho ,p) \equiv  c^2_{s}$. Moreover, a set $\{n, s, \Theta\}$ of {\em thermodynamic quantities} ({\em matter density} $n$, {\em specific entropy} $s$ and {\em temperature} $\Theta$) exists ({\em thermodynamic scheme}), which is constrained by the common thermodynamic laws \cite{Eckart, Krasinski-Plebanski, Rezzolla}. 
Namely, the conservation of matter:
\begin{equation}  
\nabla \cdot (nu) = u(n) + n \theta = 0 \, ,  \label{c-masa}
\end{equation}
where $\theta$ is the expansion of the fluid flow; 
and the {\em local thermal equilibrium relation}, which can be written as:\\[-4.5mm]
\begin{equation}
n \Theta \dif s = \dif \rho - h \dif n \, ,  \qquad h \equiv \frac{\rho+p}{n} \, , \label{re-termo}
\end{equation}
where $h$ is the {\em relativistic specific enthalpy}. Then, the {\em specific internal energy} $\epsilon$ is determined by:
\begin{equation}
\rho= n(1+\epsilon) \, .  \label{masa-energia} 
\end{equation}

When a divergence-free perfect energy tensor $T \equiv \{u, \rho, p\}$ fulfills the hydrodynamic sonic condition S we say that it defines a {\em hydrodynamic flow}. 

Another basic physical condition imposed on the thermodynamic schemes $\{n, s, \Theta\}$ is the positivity of the matter density, of the temperature and of the specific internal energy,
\begin{equation} \label{P}
\hspace{-7mm} {\rm P} :  \qquad \qquad     \Theta > 0 \, , \qquad  \quad   \rho > n > 0   \, .
\end{equation}

Finally, a coherent theory of shock waves requires the relativistic compressibility conditions  \cite{Israel, Lichnero-1, Lichnero-2}. They impose some inequalities on the derivatives of the function of state $\tau = \tau(p, s)$, $\tau = \hat{h}/n$, $\hat{h} = h/c^2$ being the dimensionless enthalpy index (note that although we set $c = 1$ in the rest of the paper, we write it explicitly here to show that $\hat{h}$ is dimensionless). In \cite{CFS-CC} we have shown that the compressibility conditions H$_1$,  $(\tau'_p)_s < 0$, $(\tau''_p)_s > 0$, only restrict the hydrodynamic quantities, and that they can be stated in terms of the function of state $c_s^2 = \chi(\rho,p)$: 
\begin{equation}
\hspace{0mm} {\rm H}_1 :  \quad 0\! <\! \chi \! <\! 1 , \ \  \  (\rho+p)(\chi \chi_{p}' + \chi_{\rho}') + 2 \chi(1-\chi) > 0    .       \label{H1-chi}
\end{equation}
However, the compressibility condition H$_2$, $(\tau'_s)_p > 0$, imposes constraints on the thermodynamic scheme and it can be stated as \cite{CFS-CC}:
\be \label{H2-Theta}
\hspace{-14mm} {\rm H}_2 : \qquad \qquad  \qquad  2 n \Theta > \frac{1}{s_{\rho}'}    \, . \qquad 
\ee
In expressions (\ref{H1-chi}) and (\ref{H2-Theta}), and hereinafter, for a function of state $f= f(\rho,p)$ we write $f'_{\rho} \equiv (\partial_{\rho}f)_p$ and $f'_p \equiv (\partial_p f)_{\rho}$.

Note that the energy conditions E, the hydrodynamic sonic condition S, and the compressibility conditions H$_1$ exclusively involve the hydrodynamic quantities $\{u, \rho, p\}$. They fully determine the hydrodynamic flow of the thermodynamic fluid in local thermal equilibrium and, consequently, restrict the admissible gravitational field as a consequence of the Einstein equations.

Instead, the positivity conditions P and the compressibility condition H$_2$ restrict the thermodynamic schemes $\{n, s, \Theta\}$ associated with a hy\-dro\-dynamic flow $\{u, \rho, p\}$. Consequently, they do not restrict the gravitational field and the admissible thermodynamics offer different physical interpretations for a given hydrodynamic perfect fluid flow.

%%%%%%%%%%%

\subsection{Isotropic radiation}
\label{subsec-isotropic_radiation}

The high level of isotropy of the cosmic microwave background radiation is usually considered as a proof that a good cosmological model must be close to a FLRW universe. This conception rests on the the Ehlers-Geren-Sachs (EGS) theorem that states \cite{EGS}: if the cosmological observer of a dust solution measures isotropic radiation, then the spacetime is a FLRW model. This result follows from a previous one by Tauber and Weinberg \cite{Tauber-EGS} on the isotropic solutions of the Liouville equation, which was later generalized for the case when an isotropic collision term exists \cite{TE-EGS}.

The general form of the Einstein equations for spacetimes with isotropic radiation measured by an irrotational observer has been obtained in \cite{FMP-isotropa}. This study shows that the geodesic character of the cosmological observer is a necessary requirement in generalizing the EGS result. In fact, Clarkson and Barrett \cite{Clarkson-1999} proved that the perfect fluid solutions with a comoving irrotational observer measuring isotropic radiation are a subclass of the thermodynamic Stephani universes, which only have a geodesic flow in the FLRW limit. 

An essential property that generates all the results in the above references is the following: a unit vector $u$ defines an observer measuring isotropic radiation if, and only if, it fulfills:
\be \label{u-CKV}
\sigma =0 ,  \qquad \dif \left[a - \frac13 \theta u\right]=0 ,
\ee
where $\sigma$, $a$ and $\theta$ are, respectively, the shear, the acceleration and the expansion of $u$.
Conditions (\ref{u-CKV}) state that $u$ is proportional to a a conformal Killing vector. Consequently, the spacetime is conformally stationary. On the other hand, the energy density, the pressure and the temperature of the radiation fluid are given by \cite{FMP-isotropa}:
\be \label{rho-p-radiacio}
\rho_r\! =\! 3 p_r\! =\! a_R \Theta_r^4, \quad \Theta_r\! =\! \Theta_0 \beta^{-1},  \quad \dif \ln \beta =  a\! -\! \frac13 \theta u .
\ee
%

%%%%%%%%%%%

\subsection{About this paper}
\label{subsec-thispaper}
%

%%%%%%%%%%

%

Although many properties on the thermodynamics of the Stephani universes are known \cite{bc-0, bc, KQS, Sussman, CF-Stephani, CFS-CC}, there are significant features that are yet to be analyzed. In Sec. \ref{sec-TSU} we obtain the speed of sound, $c_s^2= \chi(\rho,p)$, for a generic thermodynamic Stephani universe, and we outline several approaches to undertake the field equations. We also determine the corresponding associated thermodynamic schemes $\{n, s , \Theta\}$.

In order to better understand the physical meaning of the thermodynamic Stephani universes we must demand complementary significant physical qualities. In Sec. \ref{sec-qualities} we analyze how some of these constraints restrict the models. First, we impose the {\em ideal sonic condition}, $\chi=\chi(\pi) \not= \pi \equiv p/\rho$, which leads to the ideal gas Stephani universes. These models have already been studied \cite{CF-Stephani, CFS-CC} and here we summarize some results that we will use later. Second, we analyze the compatibility of the solutions with a fluid with nonvanishing thermal conductivity coefficient, and we show that only the FLRW models are possible. Third, we study the constraints on the models approaching a classical ideal gas at low temperatures. And finally, we determine the restrictions when we demand a good behavior at high temperatures.

Sec. \ref{sec-Isotropic} is devoted to studying the perfect fluids with an irrotational unit velocity measuring isotropic radiation.  Starting from the result by Clarkson and Barrett \cite{Clarkson-1999}, we obtain the constraints on the metric line element and we write it for the spherically symmetric case. We show that the {\em Dabrowski metric} \cite{Dabrowski}, which was considered in \cite{Clarkson-1999} as a cosmological model with isotropic radiation, is a solution that does not fulfill the macroscopic constraint for physical reality as a fluid in local thermal equilibrium. We also give some general properties of the ideal gas models with isotropic radiation.

In Sec. \ref{sec-Model-Nos} we study the Stephani universes modeling an ultrarelativistic gas with the comoving observer measuring isotropic radiation. They approximate a Synge gas at high temperatures and fulfill the compressibility conditions. The so-called singular model is analyzed in detail by obtaining the spacetime regions where the energy conditions hold, getting the time evolution and radial profile of the thermodynamic quantities, and studying the generalized Friedmann equation.

Finally, in Sec. \ref{sec-conclusions} we comment on the conceptual and practical interest of our results and we discuss possible future work.

%%%%%%%%%%%%%%%%%%%%%%%%%%%%%%%%%%%%%%%%%%%%%%%%%%%%%%%%%%%%
\section{Thermodynamic Stephani Universes}
\label{sec-TSU}
%%%%%%%%%%%%%%%%%%%%%%%%%%%%%%%%%%%%%%%%%%%%%%%%%%%%%%%%%%%%

In \cite{bc}, Bona and Coll showed that the Stephani universes that model the evolution of a fluid in local thermal equilibrium are those admitting a three-dimensional isometry group on two-dimensional orbits. They also showed that the metric line element of the thermodynamic Stephani universes may be written as:
\begin{subequations} \label{metrica}
\begin{eqnarray}
\dif s^2 = -\alpha^2 \dif t^2 + \Omega^2 (\dif x^2 + \dif y^2 + \dif z^2) \, ; \qquad \quad  \label{the-ste-uni}  \\
\alpha \equiv R \partial_R \ln L \, , \quad  
\Omega \equiv \frac{w}{2z} L\, , \quad L \equiv \frac{R(t)}{1+ b(t) w} \, , \quad \label{eq:termetric-1} \\[0mm] 
w \equiv \frac{2z}{1 + {\varepsilon \over 4}{\rm r}^2}   \, , \qquad r^2 \equiv x^2 + y^2 + z^2 \, , \qquad \quad \label{eq:termetric-2}
\end{eqnarray}
\end{subequations}
$R(t)$ and $b(t)$ being two arbitrary functions of time. Its symmetry group is spherical, plane or hyperbolical
depending on whether $\varepsilon$ is $1$, $0$ or $-1$. 

Furthermore, the fluid unit velocity is $u=(1/\alpha)\partial_t$, and the energy density, the pressure, the expansion and the 3-space curvature are given by
\begin{eqnarray}
\rho = \displaystyle \frac{3}{R^2} (\dot{R}^2 + \varepsilon - 4 b^2)   ,
\qquad p= - \rho - {R \over 3} {\rho'(R) \over \alpha}    ,  \qquad  \label{tdp-1}  \\[0mm]   \theta \displaystyle = \frac{3\dot{R}}{R} \not=0     ,
\quad  \qquad  \kappa \displaystyle  = \frac{1}{R^2} (\varepsilon - 4 b^2)   ,  \qquad \label{tdp-2}
\end{eqnarray}
where, for a function $f$ depending on the coordinate $t$, a dot denotes the derivative with respect to $t$. Also, if $g$ is another function of $t$, we may write $f=f(g)$ and $f'=f'(g)$. Note that the metric and the invariant quantities depend on two arbitrary functions of time $\{R(t),b(t)\}$.

The FLRW limit occurs when one of the following three equivalent conditions holds: (i) $b(t)=constant$, (ii) the fluid flow is geodesic ($\alpha = 1$), (iii) the pressure is homogeneous, $p=p(t)$.

%%%%%%%%%%%%%%%%%%%%%%%%%%%%%%%%%%%%%%%%%%%%%%%%%%%%%%%%%%%%%%%%
\subsection{Speed of sound: Indicatrix function $\chi(\rho,p)$}
\label{subsec-Chi-Stephani}
%%%%%%%%%%%%%%%%%%%%%%%%%%%%%%%%%%%%%%%%%%%%

Due to the symmetries of the metric line element (\ref{metrica}), all scalar invariants depend on two functions $(t, w)$ at most. Then, the hydrodynamic sonic condition S given in (\ref{lte-chi}) is automatically fulfilled. Now, we study the general expression of $\chi(\rho,p)$, which collects all the thermodynamic information that can be expressed using exclusively hydrodynamic quantities.

From the second equation in (\ref{tdp-1}), a direct calculation leads to:
\begin{eqnarray}
\label{eq:pi0}
\pi = \frac{p}{\rho} = \frac{a}{\alpha} -1 , \qquad  a  =  a(R) \equiv  - \frac{R\, \rho'(R)}{3 \rho} , \qquad  \\
 \chi \equiv  \frac{u(p)}{u(\rho)} = \frac{\partial_R (\rho \pi)}{\partial_R \rho}  = \pi -
 \frac{R}{3a} \partial_R \left(\frac{a}{\alpha}\right) , \qquad
\end{eqnarray}
and, from the definition of $\alpha$ in (\ref{eq:termetric-1}), it
follows
\begin{equation}
  \alpha  =  \alpha(R,w)  \equiv  \frac{1+(b-Rb')w}{1+bw}.
\label{eq:alpha}
\end{equation}
Then, from these expressions, we obtain:
\begin{eqnarray}
  \pi \! = \! \pi(R, w)\! \equiv \! \frac{a(1+bw)}{1+(b-Rb')w}-1 , \qquad \qquad \qquad 
\label{eq:pi}  \\
  \chi \!=\!  \chi(\pi,\!R)\! \equiv \!  \pi \!+\! \frac{1}{3} \!+\!
  \frac{1}{3}(\pi\!+\!1)[(\pi\!+\!1)A_1(R) \! + \! A_2(R)], \quad \quad  \label{eq:chi} \\ \label{A_i}
   A_1(R) \equiv -\frac{Rb''}{a^2b'}  , \qquad A_2(R) \equiv   \frac{Rb''}{ab'} - \frac{a'R}{a^2} - \frac{1}{a}. \qquad \quad \label{Ai(R)}
\end{eqnarray}
Thus, $\rho$ being an effective function of $R$, the functions $A_1$ and $A_2$ can be considered as depending on $\rho$, and we arrive at:
\begin{proposition} \label{prop-chi-Stephani}
The speed of sound, $c_s^2=\chi(\rho,p)$, of a thermodynamic Stephani universe {\em (\ref{metrica})} is given by:
\be
\chi(\rho, p)\! = \!  \pi \!+\! \frac{1}{3} \!+\!   \frac{1}{3}(\pi\!+\!1)[(\pi\!+\!1)A_1(\rho) \! + \! A_2(\rho)],  \quad  \label{eq:chi(rho,p)}
\ee
where $\pi= p/\rho$, and $A_1(\rho)$ and $A_2(\rho)$ are two real functions.
\end{proposition}
Every choice of these two functions determines the indicatrix function, which fixes the hydrodynamic properties of a specific thermodynamic Stephani universe.   

%%%%%%%%%%%%%%%%%%%%%%%%%%%%%%%%%%%%%%%%%%%%%%%%%%%%%%%%%%%%%%%%%%%%%%%%%%%%%%%%%%%
\subsection{On the generalized Friedmann equations}
\label{subsec-eq-Stephani}
%%%%%%%%%%%%%%%%%%%%%%%%%%%%%%%%%%%%%%%%%%%%%%%%%%%%%%%%%%%%%%%

When studying these physical properties of the solutions we can adopt different approaches. On the one hand, we can give the functions of time $R(t)$ and $b(t)$, which determine a solution, and from (\ref{tdp-1}) and (\ref{Ai(R)}), calculate the functions $\rho(R)$, $A_1(R)$ and $A_2(R)$. Then, we can obtain $R(\rho)$, and (\ref{eq:chi(rho,p)}) gives the indicatrix function $\chi(\rho,p)$, which would have to be analyzed to know the thermodynamic meaning of this specific solution. 

On the other hand, we can prescribe the functions $A_1(\rho)$ and $A_2(\rho)$ so that the indicatrix function $\chi(\rho,p)$ has specific physical properties. This choice defines a differential system for the metric functions $R(t)$ and $b(t)$, that must be solved.

This second standpoint is the one we take when studying the ideal gas Stephani universes \cite{CF-Stephani}. The ideal gas equation of state imposes the indicatrix function to be of the form $\chi = \chi(\pi)$, $\pi=p/\rho$ \cite{CFS-LTE}. Then, $A_1(\rho)$ and $A_2(\rho)$ are, necessarily, constant functions, and the study of the subsequent equations (\ref{Ai(R)}) leads to distinguish the regular and the singular models, and to obtain five possible classes of ideal gas Stephani models \cite{CF-Stephani, CFS-CC} (see subsection \ref{subsec-ideal} below). 

In studying the field equations for a given choice of the functions $A_i(\rho)$, it could be suitable to consider all the functions of $t$ as depending on the variable $\rho$. Then, Eqs. (\ref{Ai(R)}) are equivalent to:
\begin{eqnarray}
\hspace{-8mm}  a_1(\rho) R^2 -R''(\rho) R - R'(\rho)[a_2(\rho) R\! -\! 2 R'(\rho)] = 0, \label{R(rho)}
\\[1mm]
\displaystyle b(\rho) = \int \! \left[ R^2(\rho) e^{\!-\!\int \! a_2(\rho) \dif \rho}\right]  \dif \rho  ,  \label{b(rho)} \\[0mm]
a_1(\rho) \equiv \frac{A_1(\rho)}{9 \rho^2} , \qquad a_2(\rho) \equiv \frac{A_2(\rho)+3}{3 \rho} .
\end{eqnarray}
Thus, we obtain the second-order differential equation (\ref{R(rho)}) for $R(\rho)$. Once solved, expression (\ref{b(rho)}) determines $b(\rho)$. Finally, we must solve the generalized Friedmann equation for $\rho(t)$ that follows from (\ref{tdp-1}):
\be
\rho R^2(\rho)= \displaystyle 3 \,[R'(\rho)^2 \dot{\rho}^2 + \varepsilon - 4\, b(\rho)^2] . \label{gfe}
\ee
%

%%%%%%%%%%%%%%%%%%%%%%%%%%%%%%%%%%%%%%%%%%%%%%%%%%%%%%%%%%%%%%%%%%%%%%%%%%%%%%%%%%%
\subsection{Thermodynamic schemes: Entropy, matter density and temperature}
\label{subsec-Schemes-Stephani}
%%%%%%%%%%%%%%%%%%%%%%%%%%%%%%%%%%%%%%%%%%%%%%%%%%%%%%%%%%%%%%%

Each of the solutions considered above can be furnished with a family of thermodynamic schemes $\{n, s, \Theta\}$, which offer different thermodynamic interpretations of this solution. In \cite{CFS-LTE} we have shown that the specific entropies $s$ and the matter densities $n$ associated with $T$ are of the form $s = s(\bar{s})$ and $n = \bar{n}/N(\bar{s})$, where $s(\bar{s})$ and $N(\bar{s})$ are arbitrary real functions of a particular solution $\bar{s} = \bar{s}(\rho, p)$ to the equation $u(s)=0$, and $\bar{n} = \bar{n}(\rho,p)$ is a particular solution to Eq. (\ref{c-masa}).  The metric function $w$ given in (\ref{eq:termetric-2}) is a function of state that plays an important role in obtaining these thermodynamic schemes. From expression (\ref{eq:pi}) we obtain:
\begin{equation} \label{w(rhop)}
w =\frac{\pi  + 1 -a(R)}{(\pi \!+ \!1)[R b'(R)\!-\!b(R)]\! +\!a(R)b(R)}  \equiv w(\rho,p) \, .
\end{equation}
Note that $w = w(\rho,p)$ is a function of state whose dependence on $p$ is explicit, while its dependence on $\rho$ is partially implicit through the function of time $R(\rho)$. Moreover, we have that $w$ fulfills $u(w) = 0$. Consequently, the specific entropy is an arbitrary real function depending on $w$, $s = s(w)$ \cite{CFS-LTE}. 

On the other hand, from the expression (\ref{tdp-2}) of the expansion it follows that $\bar{n} = L^{-3}$ is a particular solution of the matter conservation equation (\ref{c-masa}). Then, taking into account  the expression (\ref{eq:termetric-2}) of $L$, we obtain:
\begin{proposition} \label{prop-s-n-Stephani}
The thermodynamic schemes associated with a thermodynamic Stephani universe {\em (\ref{metrica})} are determined by a specific entropy $s$ and a matter density $n$ of the form:\\[-5mm]
%
%\begin{equation}  \label{s-n-Stephani}
%s(\rho, p)\! =\! s(w)    ; \quad  n(\rho,p)\! =\! \frac{(\pi \!+\!1\! -\! a)[N(w)]^{-1}}{(\pi \! +\! 1)(Rb'\!-\!b) +a b}   ,
%\end{equation}
%
%
\begin{equation}  \label{s-n-Stephani}
s(\rho, p)\! =\! s(w)    ; \quad  n(\rho,p)\! =\! \frac{(1+b w)^3[N(w)]^{-1}}{R^3}   ,
\end{equation}
where $s(w)$ and $N(w)$ are two arbitrary real functions of the function $w=w(\rho, p)$ given in {\em (\ref{w(rhop)})}, and $b(R)$ and $R$ depend on $\rho$ through the function $R= R(\rho)$.
\end{proposition}

The temperature of the thermodynamic scheme defined by each pair $\{s, n\}$ given in the proposition above can be obtained from the thermodynamic relation (\ref{re-termo}) as: 
\begin{equation}  \label{T-Stephani-1}
\Theta = - \frac{\rho+p}{n^2}\left[\frac{\partial n}{\partial s}\right]_{\rho} =  - \frac{\rho+p}{n^2s'(w)}\left[\frac{\partial n}{\partial w}\right]_{R}  . \qquad \quad  \\  
\end{equation}
Then, taking into account expressions (\ref{eq:pi0}), (\ref{eq:alpha}) and (\ref{s-n-Stephani}), we obtain: 
\begin{proposition} \label{prop-T-Tmodels}
For a thermodynamic Stephani universe {\em (\ref{metrica})}, the temperature $\Theta$ of the thermodynamic schemes given in proposition {\em \ref{prop-s-n-Stephani}} takes the expression:
%
%\begin{equation}  \label{T-Stephani}
%\Theta = \frac{(\rho+p) R^3}{s'(w) (1+ b w)^3}[N'(w)\!-\!3 N(w) b] \equiv \Theta(\rho,p) , 
%\end{equation}
%
%
\begin{equation}  \label{T-Stephani}
%{-0,5mm}
\Theta = \frac{(\rho\!+\!p) R^3[N'(w)(1\!+\! b w)\!-\!3 N(w) b]}{s'(w) (1\!+\! b w)^4} \equiv \Theta(\rho,p) ,
\end{equation}
where $w=w(\rho, p)$ is given in {\em (\ref{w(rhop)})}, and $b(R)$ and $R$ depend on $\rho$ through the function $R= R(\rho)$.
\end{proposition}
%

%%%%%%%%%%%%%%%%%%%%%%%%%%%%%%%%%%%%%%%%%%%%%%%%%%%%%%%%%%%%%%%%%%%%%%%%%%%%%%%%%%%
\subsection{Constraints for physical reality}
\label{subsec-Phys-reality}
%%%%%%%%%%%%%%%%%%%%%%%%%%%%%%%%%%%%%%%%%%%%%%%%%%%%%%%%%%%%%%%

In studying a specific Stephani universe defined by the functions $\{R(t),b(t)\}$, we must determine the spacetime domain (set of values of the coordinates $\{t, w\}$) where the functions $\{\rho(t), p(t,w)\}$ fulfill the energy conditions E given in (\ref{e-c}). And we must impose the compressibility conditions H$_1$ given in (\ref{H1-chi}) on the indicatrix function (\ref{eq:chi(rho,p)}) within this domain. This means that the functions $A_i(\rho)$ and their derivatives will be constrained by some inequalities. 

On the other hand, the functions $\{s(w), N(w)\}$ defining a thermodynamic scheme will be constrained by the positivity conditions P given in (\ref{P}) and the compressibility condition H$_2$ given in (\ref{H2-Theta}). 

The study of all these constraints for a generic Stephani universe results too formal and useless. We delay this study for specific solutions that can be obtained under the demand of meaningful physical qualities. In this paper we study some of them.

%%%%%%%%%%%%%%%%%%%%%%%%%
\section{Imposing some significant physical qualities}
\label{sec-qualities}
%%%%%%%%%%%%%%%%%%%%%%%%%

The general expressions of the hydrodynamic and thermodynamic quantities obtained above can be useful when we particularize them in looking for thermodynamic Stephani universes that model a perfect fluid with specific physical properties. We analyze in this section some of these requirements.

%%%%%%%%%%%%%%%%%%%%%%%%%
\subsection{Ideal gas Stephani universes}
\label{subsec-ideal}
%%%%%%%%%%%%%%%%%%%%%%%%%

A notable physical property that can be required for a perfect fluid solution is that it represents the evolution of a generic ideal gas, which is defined by the equation of state $p=\tilde{k} n \Theta$. In \cite{CFS-LTE} we have showed that this fact is characterized by the {\em ideal sonic condition}:
\begin{equation} \label{chi-pi}
\hspace{5mm} {\rm S^G} :  \quad \quad  \   \chi(\rho, p) = \chi(\pi) \not= \pi \, , \qquad \pi \equiv \frac{p}{\rho}   \, . \ \ 
\end{equation}
Moreover, the associated ideal thermodynamic scheme $\{n, s, \Theta\}$ is given by \cite{CFS-LTE}:
\begin{subequations} \label{esquema-ideal}
\begin{eqnarray} 
n  = {\rho \over e(\pi)}  ,  \qquad s = \tilde{k} \ln \frac{f(\pi)}{\rho}  , \qquad \Theta  =  {\pi \over \tilde{k}} e(\pi),  
\qquad  \label{esquema-ideal-a} \\
f(\pi) \!= \! f_0 \exp\{\! \! \int \! \! \phi(\pi)d\pi\} ,  \qquad
\phi(\pi) \! \equiv \! {1 \over \chi(\pi)-\pi} . \qquad   
\label{f-pi} \\
e(\pi) \!= \!e_0 \exp\{\! \! \int \! \! \psi(\pi)d\pi \}  , \qquad   \psi 
(\pi) \! \equiv \!  \frac{\pi}{\pi\!+\!1}\phi(\pi) . \qquad \label{e-pi} 
\end{eqnarray}
\end{subequations}

For the Stephani universes, the ideal condition (\ref{chi-pi}) implies $A_i(\rho) = c_i = constant$, 
and the indicatrix function (\ref{eq:chi(rho,p)}) becomes:
\begin{subequations} \label{chi-stephani-ideal}
\begin{eqnarray} 
\chi(\pi) = \frac13 c_1 \pi^2 + \gamma \pi + \delta , \qquad \quad  \\ \gamma \equiv 1+\frac13 (2c_1+c_2), \quad  \delta \equiv  \frac13 (1+c_1+c_2) . \ \ 
\end{eqnarray}
\end{subequations}

The study of the ideal gas Stephani universes was accomplished in \cite{CF-Stephani}. Depending on the principal constants $c_i$ five classes exist: (C1) $c_1=c_2=0$, (C2) $c_1=0, c_2\not=0$, (C3) $\Delta \equiv c_2^2 - 4c_1=0$, $c_1\not=0$, (C4) $\Delta >0$, $c_1\not=0$ and (C5) $\Delta <0$.
For every class, we determined the associated ideal thermodynamic scheme (\ref{esquema-ideal}) by explicitly obtaining the generating functions $f(\pi)$ and $e(\pi)$. 

On the other hand, the study of equations (\ref{A_i}), with $A_i(\rho) = c_i$,  leads to distinguish the singular models, ($a'(R)=0$), compatible with classes C2, C3, and C4, and regular models ($a'(R)\not=0$), compatible with the five classes Cn \cite{CF-Stephani}.

Note that  the ideal thermodynamic scheme (\ref{esquema-ideal}) for a Stephani ideal model must correspond to a specific choice of the functions $s(w)$ and $N(w)$ in proposition \ref{prop-s-n-Stephani} where the schemes associated with a thermodynamic Stephani universe are given. According to (\ref{esquema-ideal}), in the ideal thermodynamic schemes, $\rho/n$ and $\rho \exp[s/\tilde{k}]$ are functions of $\pi$. Checking the compatibility of this with the results of proposition \ref{prop-s-n-Stephani}, and using some expressions in \cite{CF-Stephani} for $a(R)$ and $b(R)$ particular to each model, we obtain that the functions $s(w)$ and $N(w)$ must fulfill the differential conditions:
\begin{subequations} \label{s(w)-N(w)}
\begin{eqnarray} 
s'(w)  = \frac{1}{\sigma_0+ \sigma_1 w  + \sigma_2 w^2} ,   \\[1mm]  \frac{N'(w)}{N(w)} = \frac{\mu_0+ \mu_1 w}{\nu_0+ \nu_1 w + \nu_2 w^2} , 
\end{eqnarray}
\end{subequations}
where the constants $\sigma_1$, $\mu_i$ and $\nu_i$ depend on the parameters of the specific Stephani ideal model.

Note that, for an ideal gas, the {\em positivity conditions} P given in (\ref{P}) imply that the energy conditions E become (here we shall consider nonshift perfect fluids, $\rho \not=p$):
\begin{equation}  \label{e-c-gas}
\hspace{-8mm} {\rm E}^{\rm G} : \qquad \qquad \quad  0 < \pi < 1 \, . \qquad 
\end{equation}

On the other hand, we know \cite{CFS-CC} that, for the ideal gas solutions, the compressibility conditions H$_1$ and H$_2$ state that the indicatrix function $\chi(\pi)$ in the domain $]0,1[$ must fulfill:
\begin{eqnarray}
\hspace{-5mm} {\rm H}_1^{\rm G} \!:  \ \        0\! < \!\chi \!< \!1  , \ \   \zeta \! \equiv \! (1\!+\!\pi)(\chi\!-\!\pi) \chi'  \!+ \! 2 \chi(1\!-\!\chi)\! > \!0    ,   \quad     \label{cc-ideal-H1}  \\[2mm]
\hspace{-5mm} {\rm H}^{\rm G}_2 : \qquad \qquad    \xi \equiv (2 \pi + 1) \chi(\pi) - \pi > 0 \, . \qquad  \label{cc-ideal-H2} \ \
  \end{eqnarray}
In \cite{CFS-CC} we have analyzed when an ideal gas Stephani model fulfills the above compressibility conditions and when it has a physically reasonable behavior at low or high temperature.

%%%%%%%%%%%%%%%%%%%%%%%%%%%%%%%%%%%%%%%%%%%%%%%%%%%%%%%%%%%%%%%%%%%%%%%%%%%%%%%%%%%
\subsection{Schemes compatible with thermal conductivity}
\label{subsec-Schemes-Thermal}
%%%%%%%%%%%%%%%%%%%%%%%%%%%%%%%%%%%%%%%%%%%%%%%%%%%%%%%%%%%%%%%

According to the theory of thermodynamics of irreversible processes \cite{Eckart, Rezzolla}, the transport coefficients of thermal conductivity, of shear viscosity, and of bulk viscosity appear in the constitutive equations linking dissipative fluxes (anisotropic pressures, bulk viscous pressure and energy flux) with the kinematic coefficients of fluid flow (shear, expansion and acceleration).

A nonperfect fluid is a fluid with at least a nonzero transport coefficient. For this fluid, the energetic evolution is, generically, described by an energy tensor with energy flux and anisotropic pressures. However, when a nonperfect fluid admits particular evolutions in which the dissipative fluxes vanish, these evolutions are well described by a perfect energy tensor, and are usually called {\em equilibrium states} \cite{Rezzolla}. Moreover, all the thermodynamic relations of the perfect fluid hydrodynamics remain valid. Furthermore, the shear, the expansion and the acceleration of the fluid undergo strong restrictions as a consequence of the constitutive equations. For such equilibrium states \cite{Rezzolla}: (i) If the shear viscosity coefficient does not vanish, then the fluid shear vanishes; (ii) If the bulk viscosity coefficient does not vanish, then the fluid expansion vanishes; (iii) If the thermal conductivity coefficient does not vanish, then the fluid acceleration is constrained by the relation:
\be \label{Fourier}
a = - \perp \dif \ln \Theta \, , 
\ee
where $\perp$ denotes the orthogonal projection to the fluid velocity.

Then, under some kinematic constraints of the fluid flow, a nonperfect fluid can evolve as a perfect fluid because the dissipative fluxes can vanish, even if the transport coefficients are nonzero.  

For example, the FLRW universes can model a thermodynamic perfect fluid in isentropic evolution. Nevertheless, they could also model the evolution of a fluid with nonvanishing thermal conductivity and shear-viscosity coefficients. Indeed, in this case we have a geodesic and shear-free flow, and any homogeneous temperature is compatible with (\ref{Fourier}).      

In \cite{CFS-PSS,CFS-RSS} we have shown that the ideal Szekeres-Szafron models can be interpreted as inviscid fluids with a nonvanishing thermal conductivity coefficient. The thermodynamic Stephani universes studied here have a shear-free flow and, consequently, are compatible with a nonvanishing shear-viscosity coefficient. Now, we study if thermodynamic schemes, which are compatible with a nonvanishing thermal conductivity coefficient, exist.

A strict Stephani universe has a nonvanishing acceleration $a = \perp \dif \ln \alpha$, where $\alpha$ is given in (\ref{eq:termetric-1}). Then, the constraint (\ref{Fourier}) states $\partial_w(\alpha \Theta)=0$. If we impose this condition on the temperature (\ref{T-Stephani}) we obtain that no solution exists for a nonconstant $b(t)$. Consequently, we arrive at:
\begin{proposition} \label{prop-Stephani-conductivity}
A thermodynamic Stephani universe {\em (\ref{metrica})} can model a fluid with nonvanishing shear-viscosity. It also models a fluid with non-vanishing thermal conductivity coefficient if, and only if, it is a FLRW universe.
\end{proposition}
%

%%%%%%%%%%%%%%%%%%%%%%%%%%%%%%%%%%%%%%%%%%%%%%%%%%%%%%%%%%%%%%%%%%%%%%%%%%%%%%%%%%%
\subsection{Good behavior at low temperatures: $\chi(\rho,0)=0$}
\label{subsec-Low}
%%%%%%%%%%%%%%%%%%%%%%%%%%%%%%%%%%%%%%%%%%%%%%%%%%%%%%%%%%%%%%%

A classical ideal gas is an ideal gas, $p=\tilde{k} n \Theta$, with the internal energy proportional to the temperature, $\epsilon = c_v \Theta$. The indicatrix function of a classical ideal gas is of the form \cite{CFS-CIG}:
\be
\chi_c = \frac{\gamma \pi}{1+ \pi} = \frac{\gamma p}{\rho+ p}, \label{CIG}
\ee
where $\gamma \equiv 1 + \tilde{k}/c_v$ is the adiabatic index. 

The expression (\ref{chi-stephani-ideal}) for the indicatrix function of the ideal gas Stephani solutions shows that no strict Stephani universe exists modeling the evolution of a classical ideal gas. Only the FLRW limit enables a barotropic solution modeling a classical ideal gas in isentropic evolution (see \cite{CFS-CIG} for more details). 

Anyway, we can analyze the thermodynamic Stephani models that approach a classical ideal gas in the vicinity of $p=0$, which is the region where the classical ideal gas equation of state is a good model. 

For the indicatrix function (\ref{CIG}) of a classical ideal gas, $\chi_c(\pi) = \tilde{\chi}_c(\rho,p)$, we have:
\begin{eqnarray}
\chi_c(0) = \tilde{\chi}_c(\rho,0) = 0,  \qquad \chi_c'(0) = \gamma ,         \label{cc-ideal}  \\[1mm]
\hspace{0mm} \partial_p \tilde{\chi}_c(\rho,0) = \frac{\gamma}{\rho} , \qquad  \partial_{\rho} \tilde{\chi}_c(\rho,0) = 0   .   \end{eqnarray}
And for the indicatrix $\chi(\rho,p)$ of a generic thermodynamic Stephani universe (\ref{eq:chi(rho,p)}) we have:
\begin{subequations} \label{chi-stephani-low}
\begin{eqnarray}
\chi(\rho,0)=\frac13[1+ A_1(\rho)+ A_2(\rho)], \\[0mm]   \partial_p \chi(\rho,0) = \frac{1}{\rho}\{1 + \frac13[2A_1(\rho)+ A_2(\rho)]\} ,    \\[0mm]
 \partial_{\rho} \chi(\rho,0) =  \frac13[A_1'(\rho)+ A_2'(\rho)]    .   
\end{eqnarray}
\end{subequations}
Consequently, we obtain:
\begin{itemize}
\item[(i)] 
A thermodynamic Stephani universe approaches a classical ideal gas up to zero order at $p=0$ if, and only if, 
the functions $A_i(\rho)$ are constrained by the condition:
\be
1+ A_1(\rho)+ A_2(\rho) =0 , \label{zero-p=0}
\ee
\item[(ii)] 
A thermodynamic Stephani universe approaches a classical ideal gas (with adiabatic index $\gamma$) up to first order at $p=0$ if, and only if, it models a generic ideal gas ($A_i(\rho)=c_i)$ with an indicatrix function of the form:
\be
\chi(\pi) =  \gamma \pi + (\gamma-2/3) \pi^2 . \label{first-p=0}
\ee
\end{itemize}
As already pointed out in \cite{CFS-CC}, if $\gamma >1$ the indicatrix function (\ref{first-p=0}) fulfills the compressibility conditions H$_1^G$ and H$_2^G$ in an interval $[0, \pi_M]$.

%%%%%%%%%%%%%%%%%%%%%%%%%%%%%%%%%%%%%%%%%%%%%%%%%%%%%%%%%%%%%%%%%%%%%%%%%%%%%%%%%%%
\subsection{Good behavior at high temperatures: $\chi(\rho,\rho/3)=1/3$}
\label{subsec-High}
%%%%%%%%%%%%%%%%%%%%%%%%%%%%%%%%%%%%%%%%%%%%%%%%%%%%%%%%%%%%%%%

The macroscopic equation of state of a relativistic nondegenerate monoatomic gas (Synge gas) can be expressed by means of second kind modified Bessel functions \cite{Synge, Rezzolla}, and several simpler analytical approaches have been proposed \cite{MFS-Synge}. The Taub-Mathews equation of state \cite{Mathews, Rezzolla} approximates the Synge gas one at first order at both low and high temperatures, and its indicatrix function is given by \cite{MFS-Synge}:
\be
\chi_{_{T\!M}} = \frac{\pi (5 - 3 \pi)}{3(1+ \pi)} = \frac{p(5 \rho - 3p)}{3 \rho (\rho+ p)}. \label{Synge} 
\ee
At first order in $p=0$ the Synge gas coincides with a classical ideal gas with adiabatic index $5/3$. Thus, the behavior at low temperatures of the Synge gas has been analyzed in the above subsection. Now, we analyze the thermodynamic Stephani models that approach a Synge gas at high temperatures.

The indicatrix function (\ref{Synge}), $\chi_{_{T\!M}}(\pi) = \tilde{\chi}_{_{T\!M}}(\rho,p)$, approximates a Synge gas in the interval $[0, 1/3]$. When the temperature increases and tends to infinity $\pi$ approaches $1/3$ ($\rho = 3p$) and $\rho$ tends to infinity. We have:
\begin{eqnarray}
\chi_{_{T\!M}}\!(1/3) =\tilde{\chi}_{_{T\!M}}\!(\rho,\rho/3)\!=\! 1/3,  \quad \chi_{_{T\!M}}'(1/3) = 1/2 ,   \qquad    \label{cc-ideal}  \\[1mm]
\partial_p \tilde{\chi}_{_{T\!M}}\!(\rho,\rho/3) = \frac{1}{2\rho} , \quad  \partial_{\rho} \tilde{\chi}_{_{T\!M}}\!(\rho,\rho/3) = -\frac{1}{6\rho}   .  \qquad  
\end{eqnarray}
And for the indicatrix, $\chi(\rho,p)$, of a generic thermodynamic Stephani universe (\ref{eq:chi(rho,p)}) we have:
\begin{subequations} \label{chi-stephani-high}
\begin{eqnarray}
\chi(\rho,\rho/3)=\frac23\{1 + \frac23[\frac43 A_1(\rho)+ A_2(\rho)]\}, \qquad \ \  \\[1mm]  
\partial_p \chi(\rho,\rho/3) = \frac{1}{\rho}\{1 + \frac13[\frac83 A_1(\rho)+ A_2(\rho)]\} , \qquad \ \   \\[0mm]
 \partial_{\rho} \chi(\rho,\rho/3)\! = \!-\frac13 \partial_p \chi(\rho,\rho/3)\! +\! \frac49[\frac43 A_1'(\rho)\!+\! A_2'(\rho)]  . \qquad   \ \
\end{eqnarray}
\end{subequations}
Consequently, we obtain:
\begin{itemize}
\item[(i)] 
A thermodynamic Stephani universe approaches a Synge gas up to zero order at $\rho=3p$ if, and only if, 
the functions $A_i(\rho)$ are constrained by the condition:
\be
9+ 16 A_1(\rho)+ 12 A_2(\rho) =0 , \label{zero-3p=rho}
\ee
\item[(ii)] 
A thermodynamic Stephani universe approaches a Synge gas up to first order at $\rho=3p$ if, and only if, it models a generic ideal gas ($A_i(\rho)=c_i)$ with an indicatrix function of the form:
\be
\chi(\pi) = \frac{1}{16} [7/3 + 10\pi - 3 \pi^2]  . \label{first-3p=rho}
\ee
\end{itemize}
As already pointed out in \cite{CFS-CC}, the indicatrix function (\ref{first-3p=rho}) fulfills the compressibility conditions H$_1^G$ and H$_2^G$ in an interval $[\pi_m, 1/3]$.

%%%%%%%%%%%%%%%%%%%%%%%%%%%%%%%%%%%%%%%%%%%%%%%
\section{Universes with isotropic radiation}
\label{sec-Isotropic}
%%%%%%%%%%%%%%%%%%%%%%%%%%%%%%%%%%%%%%%%%%%%%%%

It is known \cite{Clarkson-1999} that any perfect fluid solution with a comoving irrotational observer measuring isotropic radiation is a thermodynamic Stephani universe. In order to determine the subclass with this property we must impose equations (\ref{u-CKV}) on the cosmological observer. The first condition, $\sigma=0$, is fulfilled for any Stephani universe, and now we impose the second one. 

From expression (\ref{eq:alpha}) of $\alpha$, and taking into account that $a= \partial_w(\ln \alpha)\, \dif w$, $\theta u = - 3\, \alpha \, \dif \ln R$, a straightforward calculation shows that the second condition in (\ref{u-CKV}) is equivalent to $b''(R)=0$. Moreover, the function $\beta= R \alpha$ fulfills Eq. (\ref{rho-p-radiacio}). Consequently, we can state:
\begin{proposition} \label{prop-iso-rad}
The perfect fluid solutions with an irrotational comoving observer measuring isotropic radiation are the thermodynamic Stephani universes {\em (\ref{metrica})}, with 
\begin{equation}  \label{b(R)-isotropic}
b(R) = b_1 + b_2 R ,
\end{equation}
where $b_i$ are arbitrary constants. The (test) radiation fluid has an energy density, a pressure and a temperature given by:
%
%\be \label{rho-p-radiacio-2}
%\rho_r = 3 p_r, \quad p_r = p_0 \left(\frac{R_0}{R}\right)^4 \left[1 + \frac{b_2 w R}{1+b_1 w}\right]^4 .
%\ee
%
%
\be \label{rho-p-radiacio-3}
\rho_r = 3p_r = a_R \Theta_r^4 ,  \quad \Theta_r = \Theta_0\! \left(\!\frac{R_0}{R}\!\right)\! \!\left[1\! +\! \frac{b_2 w R}{1\!+\!b_1 w}\right] .
\ee
Moreover, the indicatrix function takes the expression {\em (\ref{eq:chi}-\ref{Ai(R)})}, with $A_1(R)=0$.
\end{proposition}
Note that this isotropic radiation defines a test fluid that is comoving with the flow of the source of the field equations: a perfect fluid with energy density and pressure given in (\ref{tdp-1}).

%%%%%%%%%%%%%%%%%%%%%%%%%%%%%
\subsection{Spherical symmetry}
\label{subsec-esferica}

%%%%%%%%%%%%%%%%%%%%%%%%%%%%%%%%%%%%%%%%%%%%%%%%%%%%%%

So far, all the results apply for spherical, plane and hyperbolic symmetries. From now on, we consider some specific models that could be developed for any symmetry, but that we only analyze for the spherically symmetric case. 
	
The metric line element of a Stephani universe with spherical symmetry can be written as (see Appendix \ref{AppendixA}):
\begin{subequations} \label{metrica-ss}
\begin{eqnarray}
\dif s^2 = -\alpha^2 \dif t^2 + \Omega^2 (\dif r^2 + r^2 \dif \tilde{\Omega}^2) \, ;  \quad  \label{the-ssste-uni}  \\
\Omega \equiv \frac{R(t)}{1 + \frac14 k(t) \, r^2} \, ,   \qquad \alpha \equiv R \, \partial_{R} \ln\Omega \,  , \quad \label{eq:sstermetric-1}  
\end{eqnarray}
\end{subequations}
$R(t)$ and $k(t)$ being two arbitrary functions of time and $\dif \tilde{\Omega}^2$ the metric of the unitary sphere. The energy density, the pressure, the expansion and the 3-space curvature are given by
\begin{eqnarray}
 \label{rho-p_esf}
		\rho = \frac{3}{R^2}(\dot{R}^2 + k) \, , \qquad p = -\rho - \frac{R}{3}\frac{\partial_{R} \rho}{\alpha} \, ,  \quad \\[0mm]  
		\theta = 3 \frac{\dot{R}}{R} \neq 0 \, , \qquad \qquad  \kappa = \frac{k}{R^2} \, .   \quad \label{curvature-ss}
\end{eqnarray}

Note that expressions in subsections \ref{subsec-Chi-Stephani}, \ref{subsec-eq-Stephani} and \ref{subsec-Schemes-Stephani} also apply for this case by changing $b \rightarrow k/4$ and $\omega \rightarrow r^2$. Only in the generalized Friedmann equation (\ref{gfe}) the change must be $\varepsilon \!- \!4 b \rightarrow k$. As a consequence, the condition for the cosmological observer to measure isotropic radiation in these coordinates is $k''(R) = 0$, namely, $k(R)= k_1 + k_2 R$, accordingly with the result in \cite{Clarkson-1999}. It is worth remarking that, using the definitions given in Appendix \ref{AppendixA}, it can be seen that this condition is coherent with (\ref{b(R)-isotropic}). Then, proposition \ref{prop-iso-rad} applies by changing $b_i \rightarrow k_i/4$ and $\omega \rightarrow r^2$.

%%%%%%%%%%%%%%%%%%%%%%%%%%%%%
\subsection{Analysis of the Dabrowski solution}
\label{subsec-roin}
%%%%%%%%%%%%%%%%%%%%%%%%%%%%%%%%%%%%%%%%%%%%%%%%%%%%%%

As explained in Sec. \ref{subsec-eq-Stephani}, a possible approach to study the physical properties of a solution is to start by prescribing the functions of time $R(t)$ and $k(t)$. Now, we consider one of the solutions considered by Dabrowski \cite{Dabrowski} in studying the general properties of the local isometric embedding of the Stephani universes (hereinafter, the \textit{Dabrowski solution}):
	\begin{equation} \label{R-k_Dabrowski}
		R(t) = D_2 \, t^2 + D_1   , \quad k(t) = - 4 D_2 R + (1\!-\!D_1^2)   ,
	\end{equation}
where $D_i$ are two real parameters. 

For the Dabrowski solution (\ref{R-k_Dabrowski}) we can obtain $\rho (t)$, $p (t, r)$ and $\pi (t, r)$ from (\ref{rho-p_esf}), and then study the energy conditions (\ref{e-c-gas}). This study was done by Barrett and Clarkson in \cite{Barrett-2000}, and they concluded that the energy conditions are fulfilled for a certain range of values of the parameters $D_i$. 

Note that the Dabrowski solution (\ref{R-k_Dabrowski}) can represent a universe with a cosmological observer measuring isotropic radiation. This fact was already pointed out in \cite{Barrett-2000}, where the physical, geometrical and observational characteristics of these inhomogeneous models were analyzed in detail. 

In the conclusions of \cite{Barrett-2000} the authors claim that the Dabrowski solutions (\ref{R-k_Dabrowski}) "admit a thermodynamic interpretation" although "there is no equation of state". We want to point out and to clarify these assertions. As explained in subsection \ref{subsec-necessary-conditions}, the existence of a formal thermodynamic scheme (subject to equations of state) can be characterized in terms of the hydrodynamic quantities by the sonic condition (\ref{lte-chi}). In fact, the Dabrowski solution fulfills the ideal sonic condition (\ref{chi-pi}). Indeed, if we use (\ref{R-k_Dabrowski}) to compute $A_1(\rho)$, $A_2(\rho)$ and $\chi(\rho,p)$, we get that $A_1 = 0$, $A_2 = -3/2$ (it is an ideal gas solution) and (\ref{chi-stephani-ideal}) becomes:
\be \label{chi-Dabrowski}
\chi(\rho, p) = \chi(\pi) = (3\pi - 1)/6 . 
\ee
Consequently, the Dabrowski solution fulfills the ideal gas equation of state $p = \tilde{k} n \Theta$. But, a question arises: do these models represent the evolution of a realistic perfect fluid? 

The indicatrix function (\ref{chi-Dabrowski}) only fulfills the causal compressibility conditions, $0 < \chi < 1$, in the interval $\pi \in ]1/3, 1[$. Moreover, $\zeta(\pi) = - \frac{17}{36} + \pi - \frac34 \pi^2$ and $\xi(\pi) = - \frac16 - \frac56 \pi + \pi^2$ are negative in the whole domain $]0, 1[$. Therefore, the indicatrix function (\ref{chi-Dabrowski}) does not fulfill the compressibility conditions H$_1^{\rm G}$ and H$_2^{\rm G}$ given in (\ref{cc-ideal-H1}) and (\ref{cc-ideal-H2}). 

Consequently, the Dabrowski solution fulfills the energy conditions and can be taken into account as a cosmological model, but it cannot be interpreted as a physically admissible thermodynamic perfect fluid.

%%%%%%%%%%%%%%%%%%%%%%%%%%%%%%%%%%%%%%%%%%%%%%%%%%%%%%%%%%%%%%%%%
%
\begin{figure*}
\includegraphics[width=0.40\textwidth]{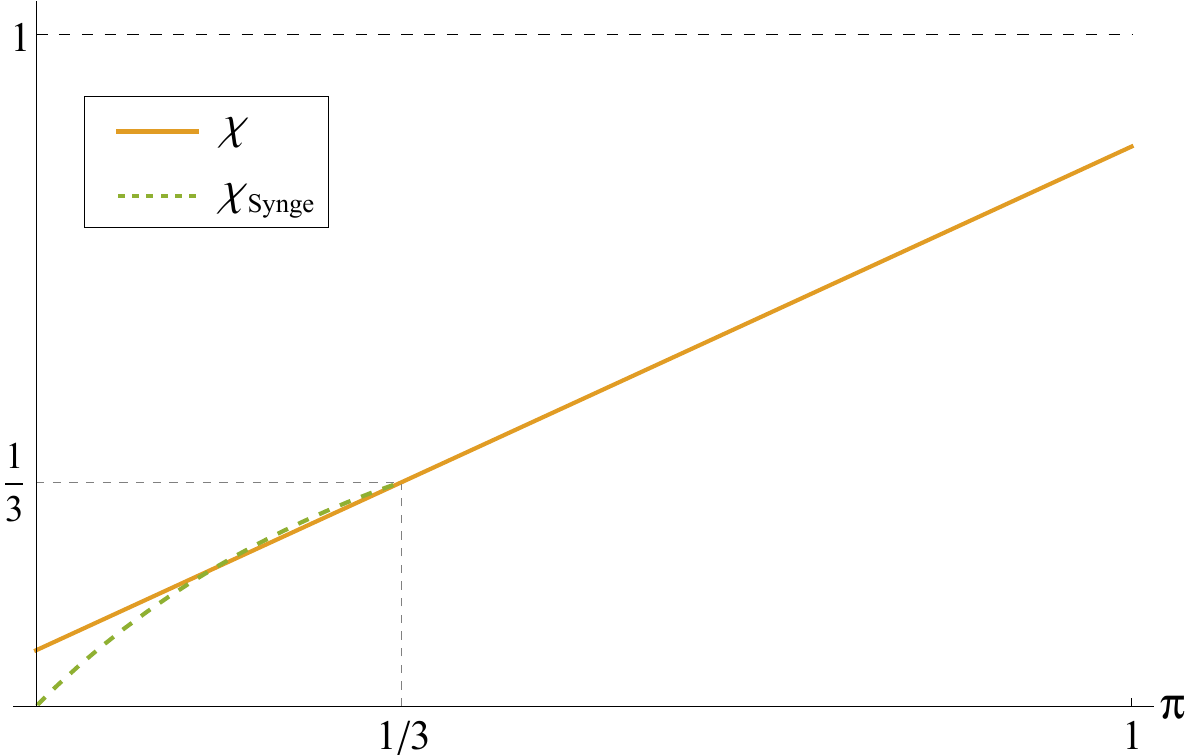} \qquad \qquad \quad
\includegraphics[width=0.40\textwidth]{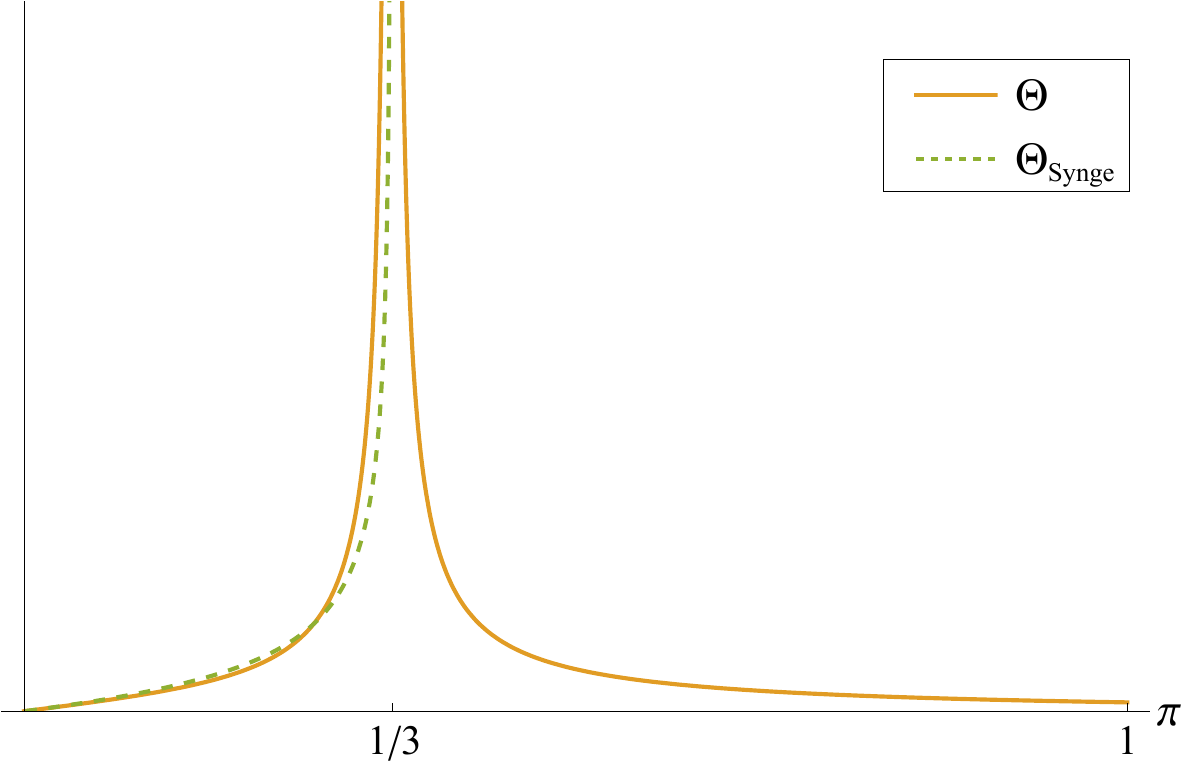}
\caption{The left panel shows the behavior of the indicatrix function $\chi(\pi)$ of our model defined in the whole interval $]0,1[$ (orange solid line), and the indicatrix function $\chi_{Synge}(\pi)$ of the Synge gas defined in the interval $]0,1/3[$ (green dashed line). The right panel shows a similar situation for the associated temperatures.}
\label{Fig-1}
\end{figure*}
%

%%%%%%%%%%%%%%%%%%%%%%%%%%%%%
\subsection{Isotropic radiation with an ideal gas source}
\label{subsec-isotropa+gasideal}
%%%%%%%%%%%%%%%%%%%%%%%%%%%%%%%%%%%%%%%%%%%%%%%%%%%%%%

In order to obtain solutions that can be interpreted as an ideal gas source with the comoving observer measuring isotropic radiation, instead of prescribing the functions $R(t)$ and $k(t)$, we will impose these physical properties to our solution and then we will study how they restrict the metric functions.

Now, all the results in subsection \ref{subsec-ideal} apply with the additional constraint $A_1= c_1 = 0$ imposed by the isotropic radiation condition (see proposition \ref{prop-iso-rad}).

With the condition $c_1=0$, the indicatrix function (\ref{chi-stephani-ideal}) of the ideal models becomes:
	\begin{equation} \label{indicatriu-isotropicradiation}
		\chi (\pi) = \gamma \, \pi + \gamma - \frac23 \, , \qquad \gamma = 1 + \frac13 \, c_2 \, .
	\end{equation}
If $1/3 < \gamma < 5/3$, this indicatrix function verifies the causal condition, $0 < \chi < 1$, for $\pi \in \, ] \, \pi_m , \pi_M \, [ \,$, where $\pi_m \equiv \frac{2}{3\gamma} - 1$ and $\pi_M \equiv \frac{5}{3\gamma} - 1$. If $2/3 \leqslant \gamma \leqslant 5/6$, these conditions are fulfilled in the whole domain $\pi  \in \, ] \, 0 , 1 \, [ \, $. 

In order to study the rest of the compressibility conditions, we need to analyze the domains in which $\zeta(\pi)$ and $\xi(\pi)$, defined in (\ref{cc-ideal-H1}-\ref{cc-ideal-H2}), are positive. Using (\ref{indicatriu-isotropicradiation}), we get $\zeta(\pi) = -\gamma(\gamma + 1)\pi^2 + \gamma \, (3 - 2\gamma) \, \pi - \gamma \, (\gamma - 4) - \frac{20}{9}$ and $\xi(\pi) = 2\gamma \, \pi^2 + \left( 3\gamma - \frac73 \right) \, \pi + \gamma - \frac23$. 

If $16/29 < \gamma < 10/3$, there exists an interval, $\pi \in \, ] \, \pi_- , \pi_+ \, [,  $ in which $\zeta(\pi)$ is positive; and
%where
%
%	\begin{equation}
%		\pi_\pm \equiv \frac{3\gamma(3 - 2\gamma) \pm \sqrt{5\gamma(29\gamma - 16)}}{6\gamma(\gamma + 1)} \, . ???????
%	\end{equation}
%
if $2/3 < \gamma < 5/6$, $\zeta(\pi)$ is positive in the whole domain $\pi  \in \, ] \, 0 , 1 \, [ \, $. 

If $\gamma > 1/2$, there exists an interval $\pi \in \, ] \, \bar{\pi}_+ , 1 \, [$ in which $\xi(\pi)$ is positive, and if $\gamma > 2/3$, it is also positive in an interval $\pi \in \, ] \, 0 , \bar{\pi}_- \, [ \, $; 
%
%, where
%
%	\begin{equation}
%		\bar{\pi}\pm \equiv \frac{7 - 9\gamma \pm \sqrt{9\gamma^2 - 78\gamma + 49}}{12\gamma} \, . ????
%	\end{equation}
%
and if $\gamma > (13 - 2\sqrt{30})/3 = \gamma_{\xi} \approx 0.682$, $\xi(\pi)$ is positive in the whole domain $\pi  \in \, ] \, 0 , 1 \, [ \, $. 

Taking all this analysis into account, we get:
\begin{proposition} \label{prop-chi-ideal-radiacio}
The indicatrix function of the ideal gas models with the comoving observer measuring isotropic radiation takes the expression {\em (\ref{indicatriu-isotropicradiation})}, and it fulfills all the compressibility conditions {\rm H}$_1^{\rm G}$ and {\rm H}$_2^{\rm G}$ in the whole domain $\pi  \in \, ] \, 0 , 1 \, [ \, $ for $\gamma_{\xi} < \gamma < 5/6$, $\gamma_{\xi}=(13 - 2\sqrt{30})/3$.
\end{proposition}

In \cite{CF-Stephani} we explained how to integrate Eqs. (\ref{Ai(R)}) for ideal gas models.
The isotropic radiation condition $c_1=0$ is only compatible with regular models of class C1 ($c_2=0$) and both singular and regular models of class C2 ($c_2 \not=0$) (see subsection \ref{subsec-ideal}). Then, a direct application of the results in \cite{CF-Stephani} leads to:
\begin{proposition} \label{prop-model-ideal-radiacio}
The generalized Friedmann equation of the ideal gas models with the comoving observer measuring isotropic radiation takes the expression:
\be \label{Friedmann-ideal-radiacio}
\rho(R) = \frac{3}{R^2}(\dot{R}^2 + k_1+ k_2 R) ,
\ee
where $\rho(R)$ depends on three different models:
\begin{itemize}
\item[i)]
{\em C2} singular ($\gamma \not=1$): $\ \rho(R) = \rho_0 (\frac{R_0}{R })^{\frac{1}{1-\gamma}}$.
\item[ii)]
{\em C2} regular ($\gamma\not=1$): $\ \rho(R) = \rho _0 (1 +  \frac{\tilde{R}_0}{R })^{\frac{1}{1-\gamma}}$.
\item[iii)]
{\em C1} regular ($\gamma=1$): $\ \rho(R) = \rho _0 \exp (\frac{\hat{R}_0}{R })$
\end{itemize}
\end{proposition}
Now, we could analyze the energy conditions (\ref{e-c-gas}) for these three cases separately, but we will leave that study for particular cases with extra physical restrictions.

%%%%%%%%%%%%%%%%%%%%%%%%%%%%%%%%%%%%%%%%%%%%%%%
\section{Ultrarelativistic gas with isotropic radiation}
\label{sec-Model-Nos}
%%%%%%%%%%%%%%%%%%%%%%%%%%%%%%%%%%%%%%%%%%%%%%%

Another possible situation of physical interest is that in which the source of the gravitational field is an ultrarelativistic fluid and the cosmological observer measures isotropic radiation. The conditions for a thermodynamic Stephani universe to behave as an ultrarelativistic fluid up to first order are studied in subsection \ref{subsec-High}. However, if we also want it to be compatible with isotropic radiation, we can only impose the good behavior at high temperatures up to zero order. 

By imposing (\ref{zero-3p=rho}) and the isotropic radiation condition $k(R) = k_1 + k_2 R$, we get that $A_1(R) = c_1 = 0$ and $A_2(R) = c_2 = -3/4$. Thus, this situation is a particular case of the one studied in the last subsection \ref{subsec-isotropa+gasideal}, with $c_2 = -3/4$. Now, $\gamma = 3/4 \in ]\gamma_{\xi}, 5/6[$, and the indicatrix function fulfills the compressibility conditions as a consequence of proposition \ref{prop-chi-ideal-radiacio}. 

Moreover, we can determine the ideal thermodynamic scheme $\{n, s , \Theta\}$ by using (\ref{esquema-ideal}). Now, the generating functions (\ref{f-pi}) and (\ref{e-pi}) take the expression:
\be \label{f-e-Synge-radiacio}
f(\pi)= \frac{f_0}{(1\!-\!3 \pi )^4} , \qquad  e(\pi)= \frac{e_0}{|1\!-\!3 \pi |(\pi \!+\!1)^3}  .
\ee
If we also take into account proposition \ref{prop-chi-ideal-radiacio} we can state:
\begin{proposition} \label{prop-Synge-radiacio}
The indicatrix function of the models approximating the Synge gas at high temperatures and with the comoving observer measuring isotropic radiation is
\be \label{indicatriu-Synge-radiation}
\chi(\rho,p)=\chi(\pi) = (9 \pi + 1)/12 .
\ee
This indicatrix function fulfills all the compressibility conditions {\em H}$_1^{\rm G}$ and {\em H}$_2^{\rm G}$ on the spacetime domain where the energy conditions, $0 < \pi <1$, hold.

The associated ideal thermodynamic scheme $\{n, s , \Theta\}$ is defined by {\em (\ref{esquema-ideal-a})}, where the generating functions $f(\pi)$ and $e(\pi)$ are given in {\em (\ref{f-e-Synge-radiacio})}. 
\end{proposition}

On the other hand, if we make $\gamma= 3/4$, proposition \ref{prop-model-ideal-radiacio} becomes:
\begin{proposition} \label{prop-model-Synge-radiacio}
The generalized Friedmann equation of the models approximating the Synge gas at high temperatures and with the comoving observer measuring isotropic radiation takes the expression {\em (\ref{Friedmann-ideal-radiacio})}, where $\rho(R)$ depends on two different models:
\begin{eqnarray} \label{singular-rho}
\hspace{-8mm} {\rm (i) \ Singular \ models:} \qquad   \ \rho(R) = \rho_0 \left(\frac{R_0}{R}\right)^4, \quad \, \\
\hspace{-8mm} {\rm (ii)\ Regular \ models:}  \qquad \rho(R) = \rho _2 \left(1+\frac{\tilde{R}_0}{R }\right)^4. \label{regular-rho}
\end{eqnarray}
\end{proposition}
\begin{figure*}
%\centerline{
\qquad \qquad \quad  $k_2< 0$  
\qquad \qquad   \quad \qquad  \qquad \qquad  \qquad \qquad  \qquad \qquad  \qquad \qquad $k_2 >0$ \\[-3mm]
%\centerline{
\parbox[c]{0.10\textwidth}{$k_1 \geq 0$} \   \parbox[c]{0.43\textwidth}{ \includegraphics[width=0.43\textwidth]{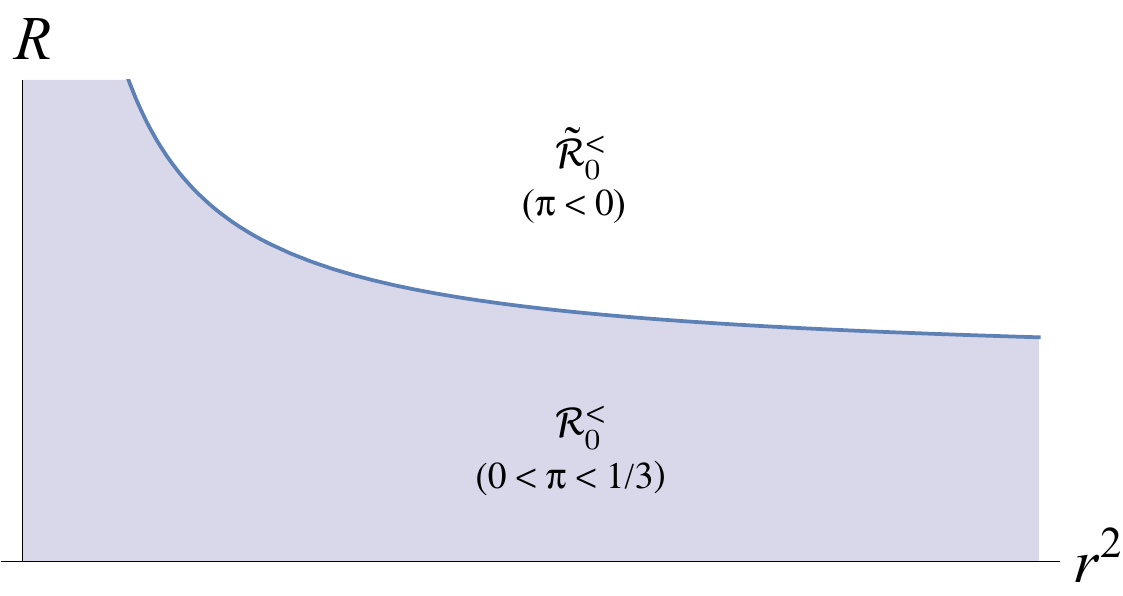}}  \  \ \parbox[c]{0.43\textwidth}{\includegraphics[width=0.43\textwidth]{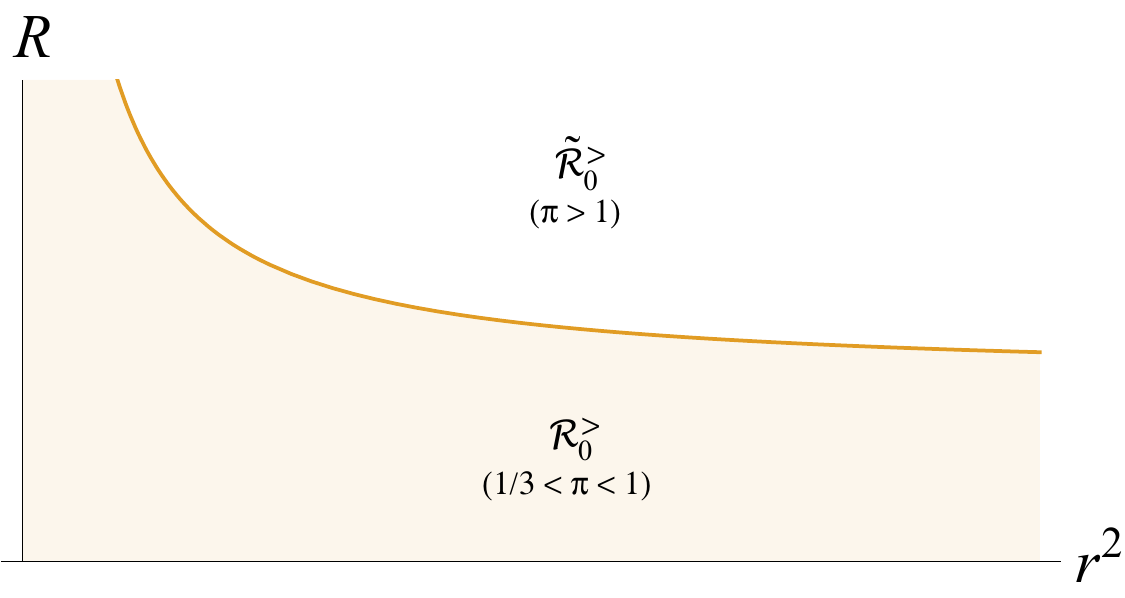}} \\
%\centerline{
\parbox[c]{0.10\textwidth}{$k_1 < 0$}\  \parbox[c]{0.43\textwidth}{\includegraphics[width=0.43\textwidth]{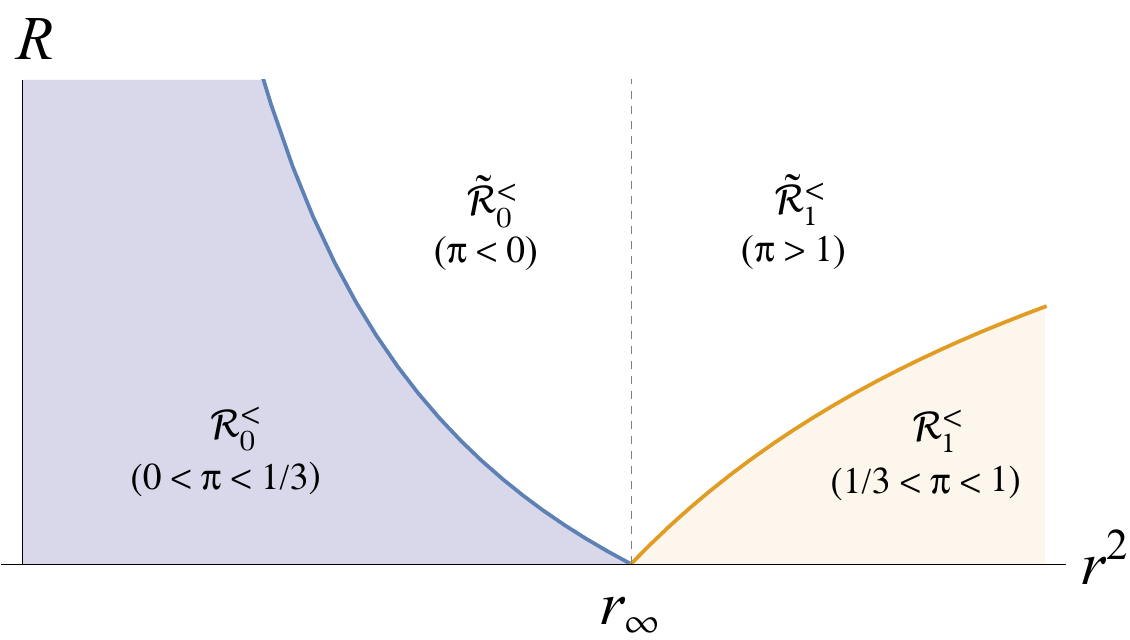}} \ \
\parbox[c]{0.43\textwidth}{\includegraphics[width=0.43\textwidth]{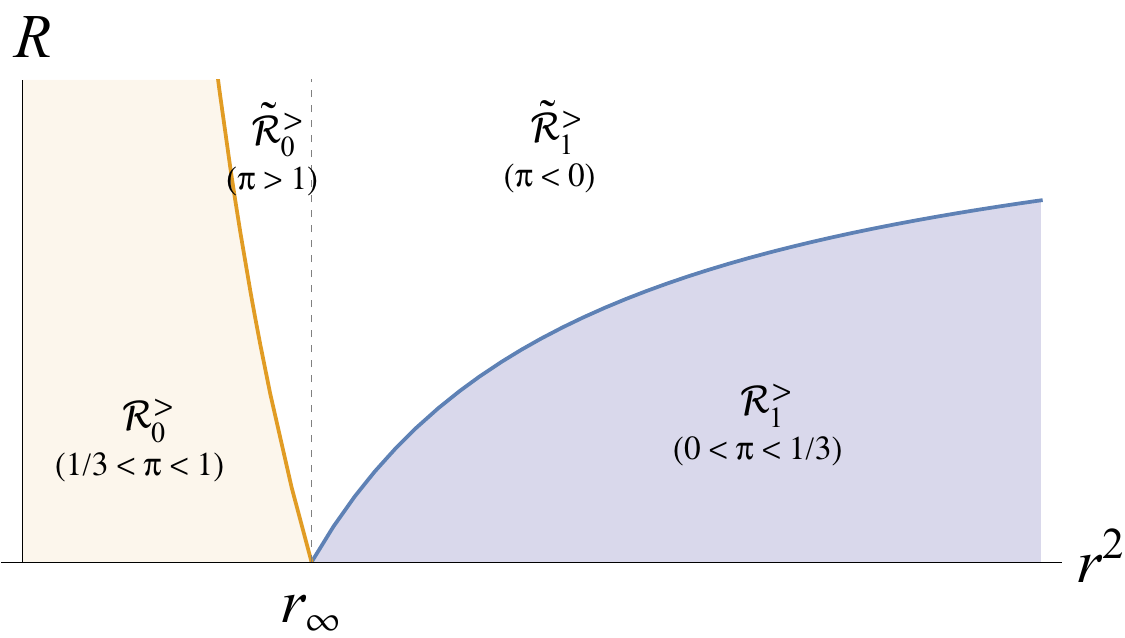}}
\caption{Spacetime coordinate domains and their physically realistic regions depending on the values of the parameters $k_i$. Upper panels: the case $k_1 \geq 0$ has a single coordinate domain ${\cal D}_0^+$. Lower pannels: the case $k_1 < 0$ has two coordinate domains, ${\cal D}_0^-$ and ${\cal D}_1^-$, separated by the straight line $r= r_{\infty}$. The dark blue lines are the hyperbolas $\pi=0$, and the light brown lines are the hyperbolas $\pi=1$. In the shaded regions the energy conditions hold, and in the dark blue regions the model approximates a Synge gas.} 
\label{Fig-2}
\end{figure*}

\Remarks 
(i) The hydrodynamic properties of the thermodynamics are given by the indicatrix function (\ref{indicatriu-Synge-radiation}), which takes the same expression for both the singular and regular models, and which fulfills the compressibility conditions in $]0,1[$. Nevertheless, this indicatrix function can only approximate the Synge one, $\chi_{Synge}$, in the interval $]0, 1/3[$ where this last one is defined (see left panel in Fig. \ref{Fig-1}). Note that this approximation gets worse the closer we get to $\pi = 0$. 

(ii) Similarly, the ideal thermodynamic scheme determined by functions (\ref{f-e-Synge-radiacio}) is defined in the interval $]0,1[$, but it only approximates the Synge thermodynamic quantities in $]0, 1/3[$. The right panel of Fig. \ref{Fig-1} shows the behavior of the temperatures.  

(iii)  In any case, the indicatrix function (\ref{indicatriu-Synge-radiation}) could be furnished with another (nonideal) thermodynamic scheme, determined by a pair of functions $\{s(r^2), n(r^2)\}$ (see subsection \ref{subsec-Schemes-Stephani}), different from (\ref{s(w)-N(w)}), which can be defined in the interval $]0,1[$. In this case, we could be modeling a physically realistic fluid but that does not satisfy the ideal gas equation of state.

(iv) The energy density $\rho_r(R,r)$, the pressure $p_r(R,r)$ and the temperature $\Theta_r(R,r)$ of the (test) radiation fluid take expressions (\ref{rho-p-radiacio-3}) (with the change $b_i \rightarrow k_i/4$) for both the singular and the regular models. Of course, the coordinate function $R(t)$ does depend on the model.

(v) The ideal thermodynamic scheme defined by the functions (\ref{f-e-Synge-radiacio}) depends on three parameters, $f_0$, $e_0$ and $\tilde{k}$. The first one, $f_0$, fixes the origin of entropy, and we can consider that the different values correspond to a sole ideal gas. The second parameter, $e_0$, modifies the specific energy in a constant factor and, consequently, the temperature and the specific volume $1/n$ change in the same factor. Be aware that $e_0$ settles the origin of internal energy. If we impose $\epsilon =0$ at zero pressure, we must take $e_0=1$. Finally, the third one, $\tilde{k}= k_B/m$ determines the mass of the gas particles. 

(vi) Singular models depend on four parameters $\{k_1, k_2, \rho_0, R_0\}$. The $k_i$ determine the function $k(R)$; $R_0$ is an initial condition for the generalized Friedmann equation (\ref{Friedmann-ideal-radiacio}), $R(t_0)= R_0$; and $\rho_0$ is the energy density at this initial time, $\rho_0 = \rho(t_0)$. However, regular models also depend on a fifth parameter $\tilde{R}_0$. The constant $\rho_2$ takes the expression $\rho_2 \equiv \rho_0 (1 + \tilde{R}_0/R_0)^{-4}$. 

From now on, we focus on the singular models (SM).

\begin{figure*}
\centerline{
\parbox[c]{0.32\textwidth}{\includegraphics[width=0.315\textwidth]{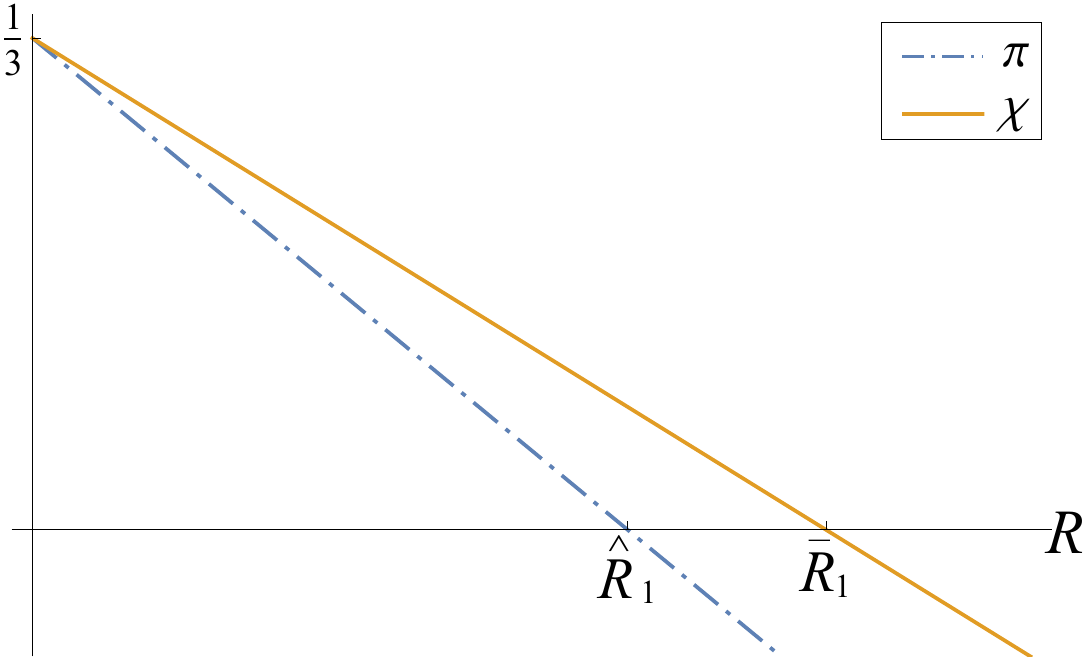}\\(a)} \ \
\parbox[c]{0.32\textwidth}{\includegraphics[width=0.315\textwidth]{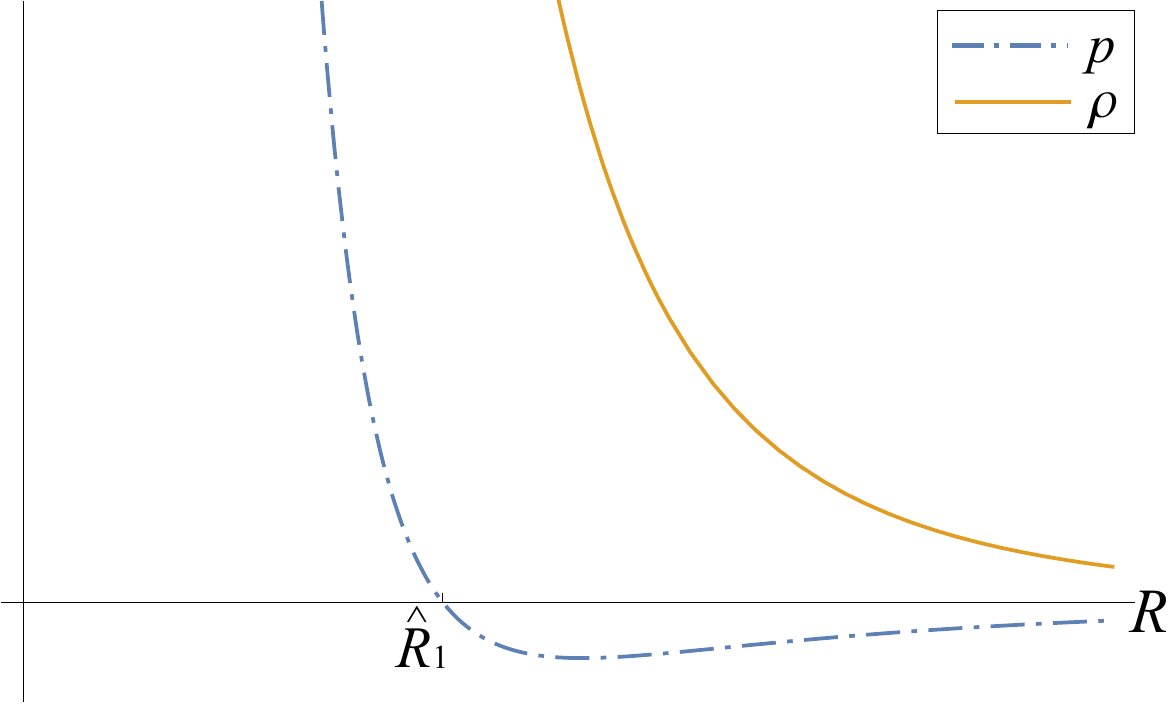}\\(b)} \ \
\parbox[c]{0.32\textwidth}{\includegraphics[width=0.315\textwidth]{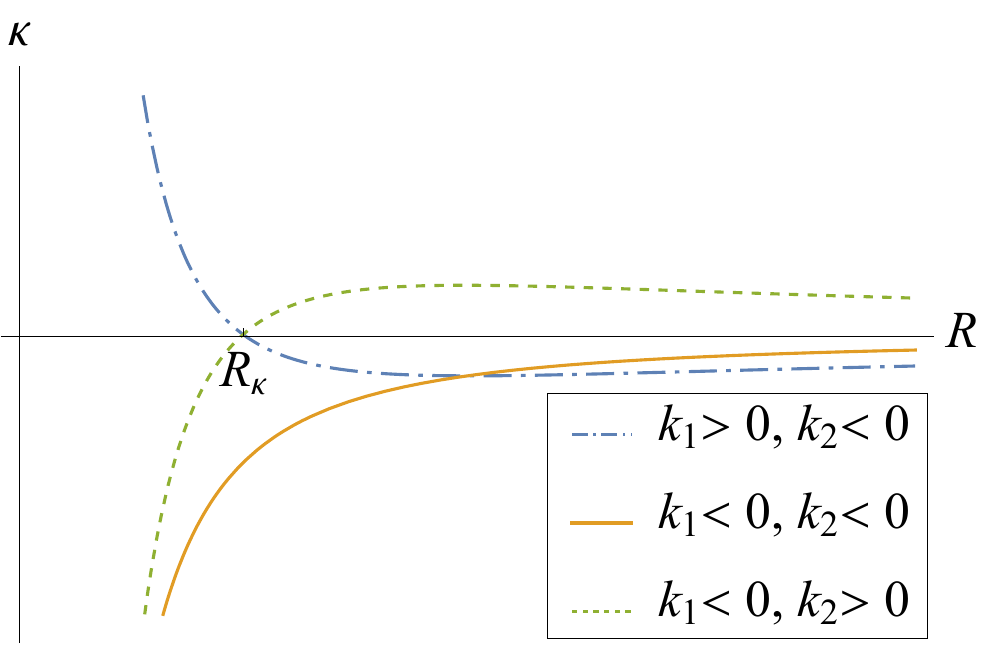}\\(c)}} 
\caption{Behavior of the models for a fixed $r_1$. (a) Evolution of the hydrodynamic quantities $\pi$ and $\chi$. (b) Evolution of the energy density and pressure. (c) Evolution of the 3-space curvature.} 
\label{Fig-3}
\end{figure*}

\begin{figure*}[hbt]
\centerline{
\parbox[c]{0.32\textwidth}{\includegraphics[width=0.315\textwidth]{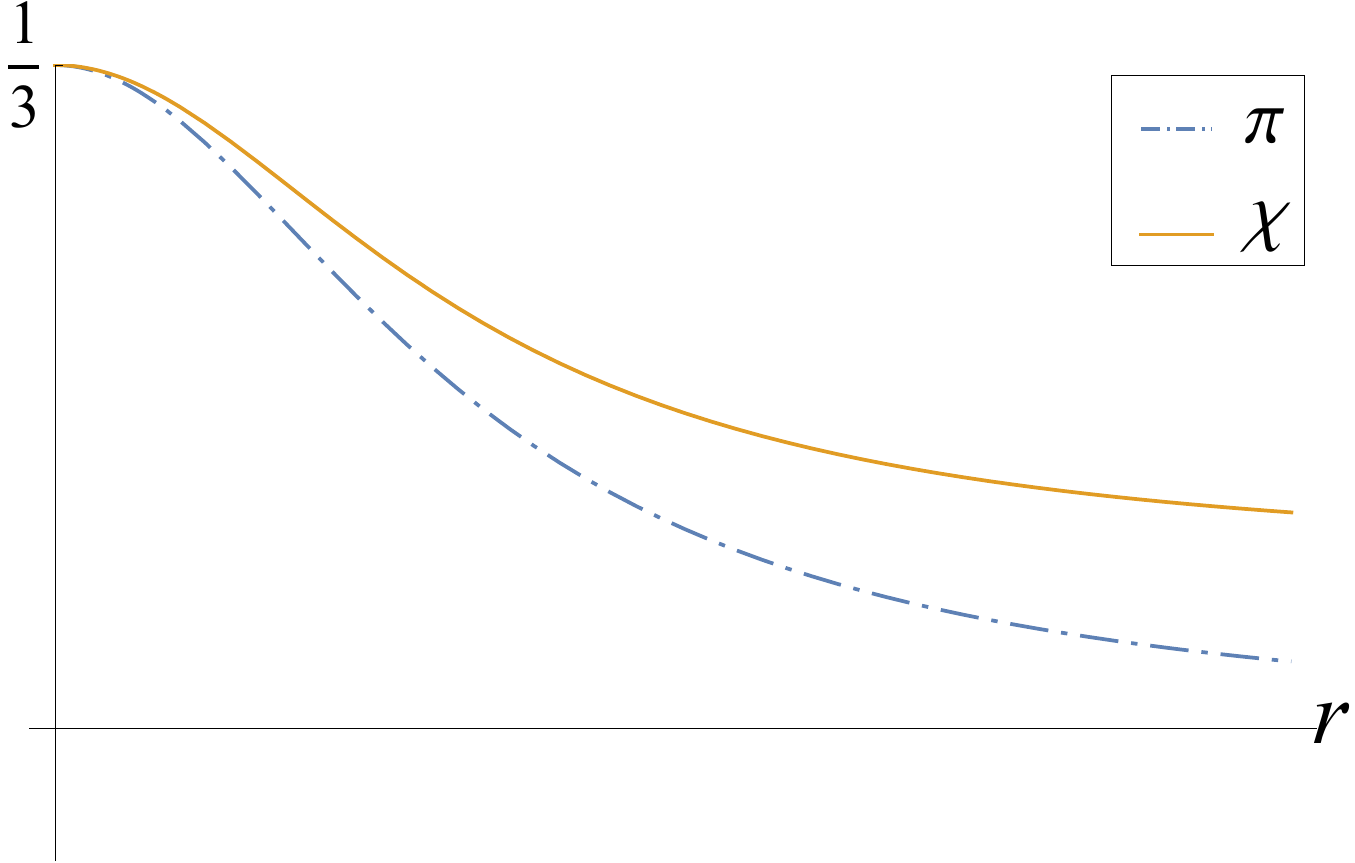}\\(a)} \ \
\parbox[c]{0.32\textwidth}{\includegraphics[width=0.315\textwidth]{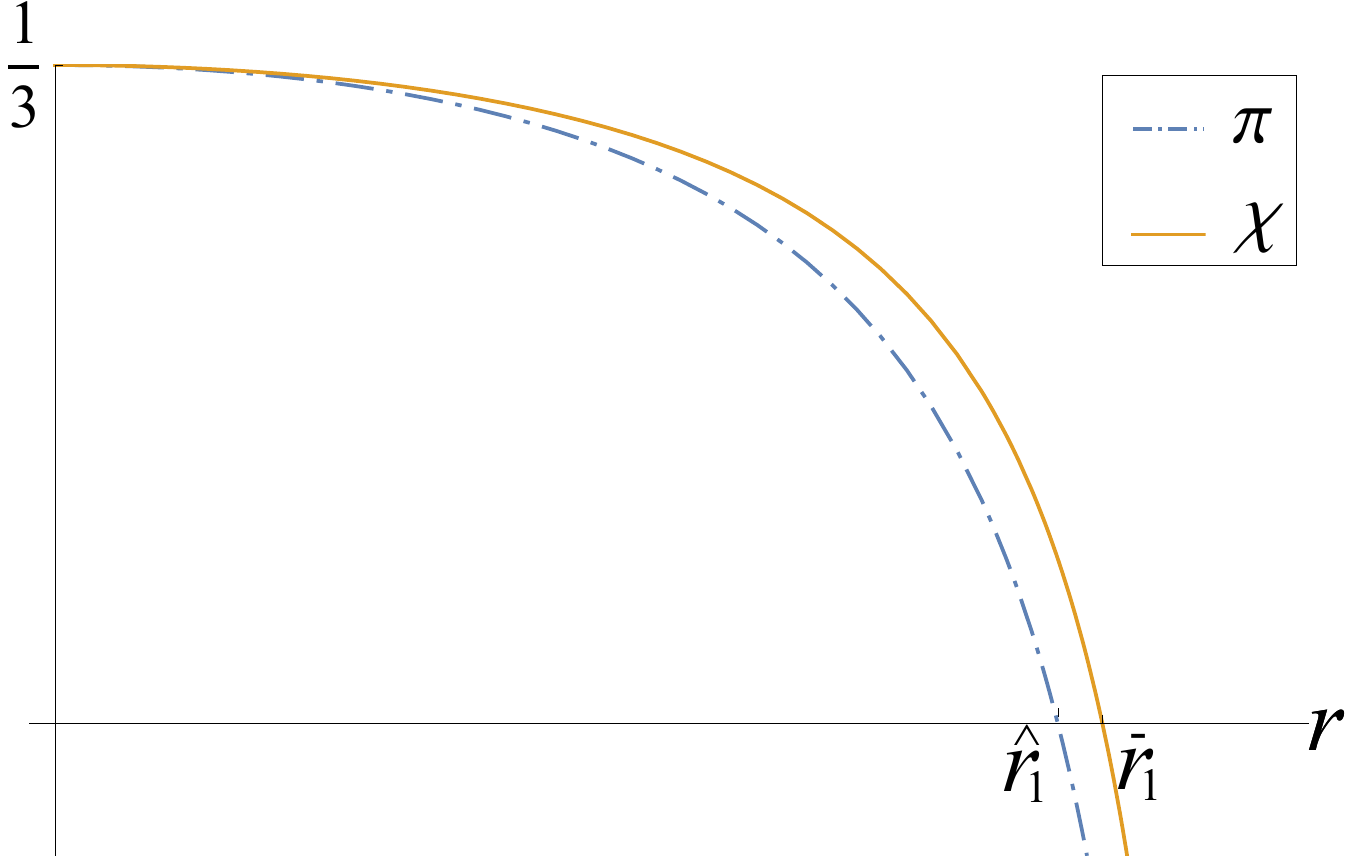}\\(b)} \ \
\parbox[c]{0.32\textwidth}{\includegraphics[width=0.315\textwidth]{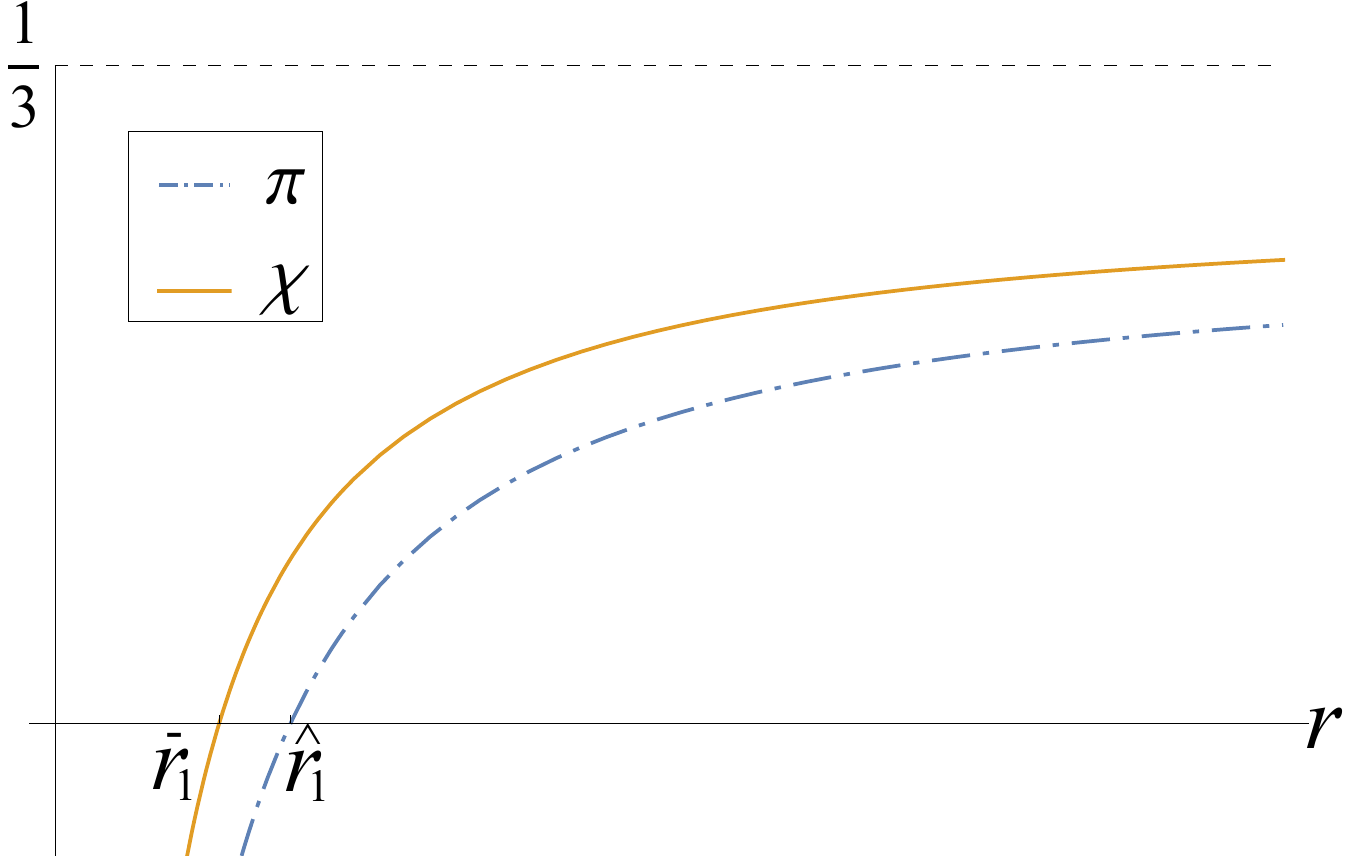}\\(c)}} 
\caption{Radial profile of the hydrodynamic quantities $\pi$ and $\chi$ depending on the values of the parameters $k_i$ for a fixed $R_1$.  (a) Case I with $R_1 \leq - k_1/(4 k_2)$. (b) Case II and case I with $R_1 > - k_1/(4 k_2)$. (c) Case III.} 
\label{Fig-4}
\end{figure*}
%

%%%%%%%%%%%%%%%%%%%%%%%%%%%%%
\subsection{SM: Spacetime domains and energy conditions}
\label{subsec-Model-Nos-condition}
%%%%%%%%%%%%%%%%%%%%%%%%%%%%%%%%%%%%%%%%%%%%%%%%%%%%%%

For the singular model the energy density is given in (\ref{singular-rho}) and the pressure takes the expression:
\be
p(R,r) = \frac13 \rho_0 \left(\frac{R_0}{R }\right)^4 \left[1 + \frac{k_2 R \, r^2}{1+\frac14 k_1 r^2}\right] .
\ee
Consequently, the hydrodynamic quantity $\pi = p/\rho$ is:
\be \label{pi-singular}
\pi(R,r) = \frac13 \left[1 + \frac{k_2 R \, r^2}{1+ \frac14 k_1 r^2}\right] .
\ee

Note that the full line element of the 3-spaces $t= $ constant vanishes at $R=0$, and
we have a big bang singularity, where both energy density and pressure diverge. However, $\pi$ takes the value $1/3$. On the other hand, when $k_1 <0$ we have another curvature singularity at $r_{\infty} = 2\sqrt{-1/k_1}$, where the pressure (and $\pi$) diverges. 

Then, when $k_1 \geq 0$ we have a single coordinate domain:
\be \label{domain}
{\cal D}_0^+ = \{(R,r), \quad R>0, \ r \geq 0\} ,
\ee
and when $k_1 < 0$ we have two coordinate domains:
\begin{subequations} \label{domains}
\begin{eqnarray}
{\cal D}_0^- = \{(R,r), \quad R>0, \ 0 \leq r < r_{\infty}\}, \ \  \\[2mm]  
{\cal D}_1^- = \{(R,r), \quad R>0, \ r > r_{\infty}\}.  \ \  
\end{eqnarray}
\end{subequations}

Now, expression (\ref{pi-singular}) enables us to analyze the regions of the different domains where the energy conditions ($0<\pi<1$) hold, and the regions where the model approximates a Synge gas ($0<\pi<1/3$). We represent all these regions in a $\{r^2, R\}$ diagram (see Fig. \ref{Fig-2}).

Note that $\pi=1/3$ if $r=0$. The spacetime events where $\pi=0$ or $\pi=1$ are defined, respectively, by the hyperbolas 
\be
R= - \frac{1}{k_2}\left[\frac{k_1}{4} + \frac{1}{r^2}\right], \quad  R= \frac{2}{k_2}\left[\frac{k_1}{4} + \frac{1}{r^2}\right]  .
\ee
Each of the domains (\ref{domain}, \ref{domains}) contains one of these hyperbolas that divides it into two regions, and the energy conditions only meet in the region next to the coordinate axes (see Fig. \ref{Fig-2}):
\begin{itemize}
\item[(i)]
Case $k_2 <0$: domains ${\cal D}_0^+$ and ${\cal D}_0^-$ contain a region ${\cal R}_0^{<}$ where the model approximates a Synge gas, $0 < \pi < 1/3$; and a region $\tilde{\cal R}_0^{<}$ where $\pi <0$. And domain ${\cal D}_1^-$ contains a region ${\cal R}_1^{<}$ where the model meets the energy conditions but it does not approximate a Synge gas, $1/3 < \pi < 1$; and a region $\tilde{\cal R}_1^{<}$ where $\pi > 1$ (see the left panels of Fig. \ref{Fig-2}).
\item[(ii)] 
Case $k_2 >0$: domains ${\cal D}_0^+$ and ${\cal D}_0^-$ contain a region ${\cal R}_0^{>}$ where the model meets the energy conditions but it does not approximate a Synge gas, $1/3 < \pi < 1$; and a region $\tilde{\cal R}_0^{>}$ where $\pi >1$. And domain ${\cal D}_1^-$ contains a region ${\cal R}_1^{>}$ where the model approximates a Synge gas, $0 < \pi < 1/3$; and a region $\tilde{\cal R}_1^{>}$ where $\pi <0$ (see the right panels of Fig. \ref{Fig-2}).
\end{itemize}
\begin{figure*}[htb]
\centerline{
\parbox[c]{0.32\textwidth}{\includegraphics[width=0.315\textwidth]{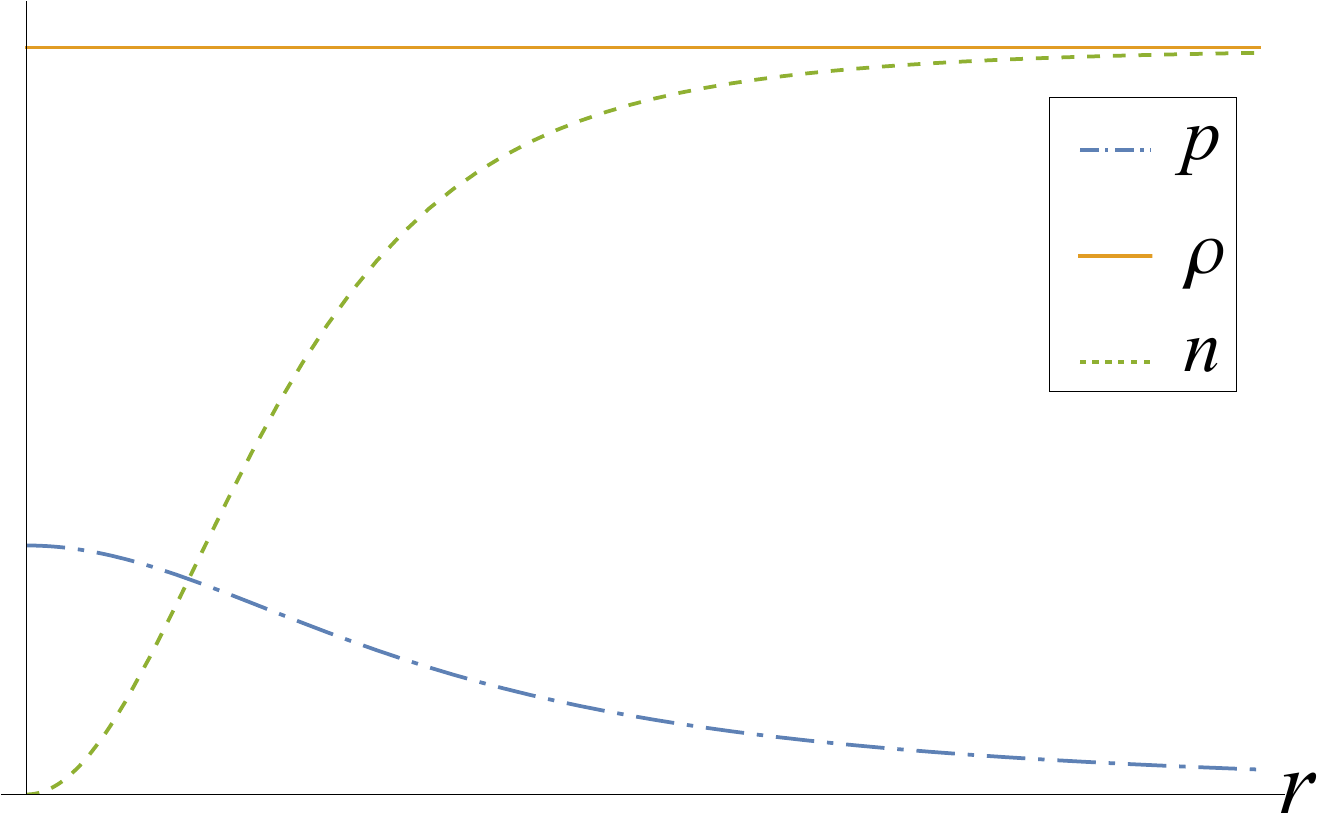}\\(a)} \ \
\parbox[c]{0.32\textwidth}{\includegraphics[width=0.315\textwidth]{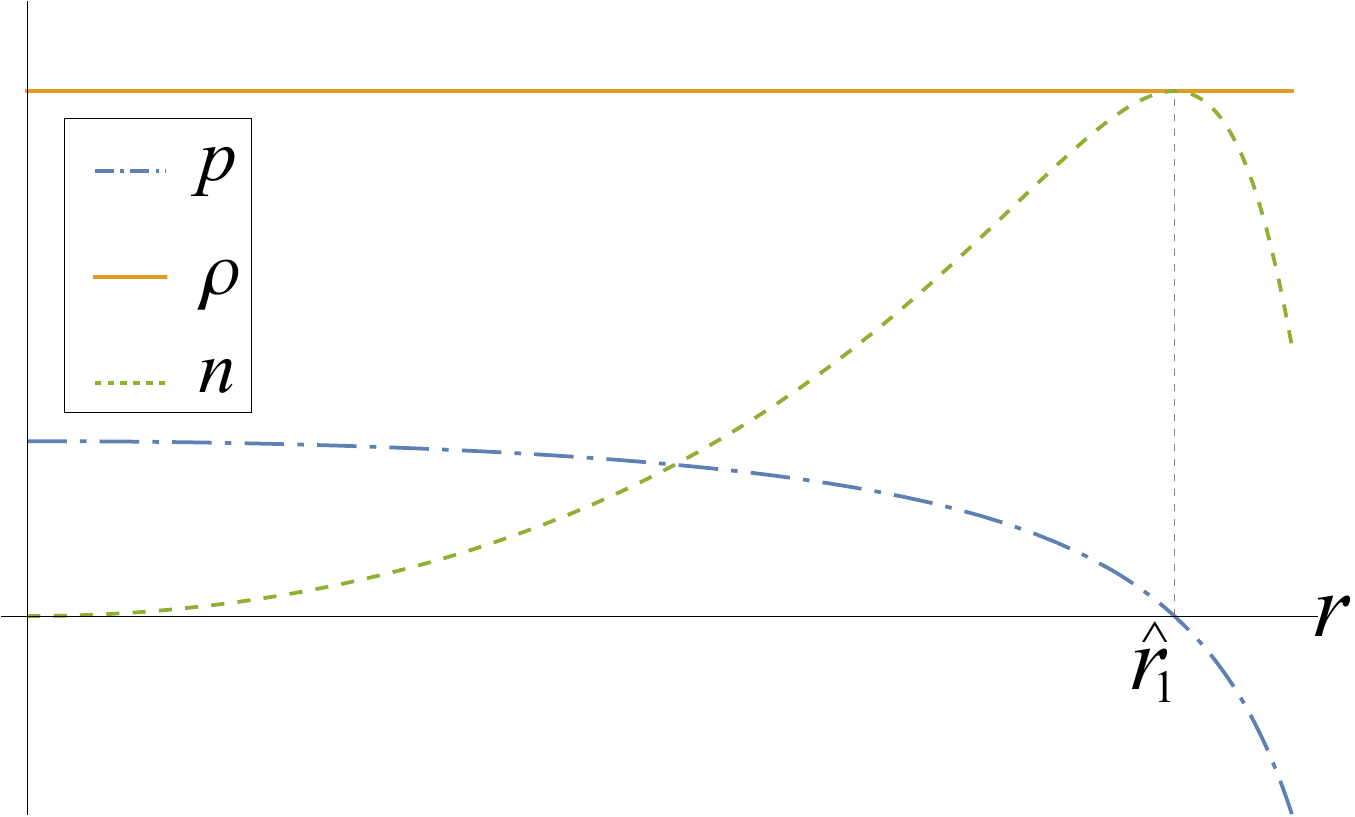}\\(b)} \ \
\parbox[c]{0.32\textwidth}{\includegraphics[width=0.315\textwidth]{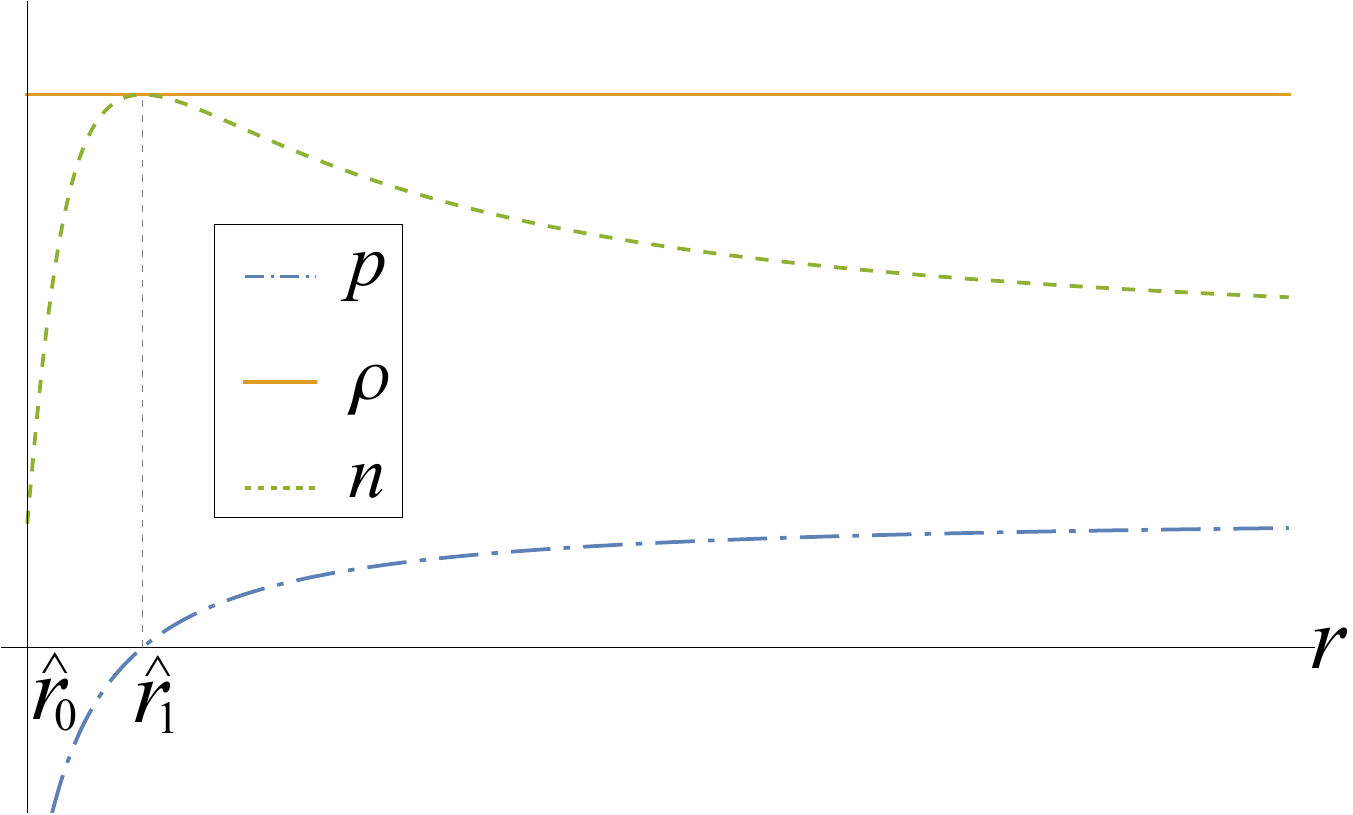}\\(c)}} 
\caption{Radial profile of the energy density $\rho$, the pressure $p$ and the matter density $n$ depending on the values of the parameters $k_i$ for a fixed $R_1$.  (a) Case I with $R_1 \leq - k_1/(4 k_2)$. (b) Case II and case I with $R_1 > - k_1/(4 k_2)$. (c) Case III.} 
\label{Fig-5}
\end{figure*}
\begin{figure*}
\centerline{
\parbox[c]{0.32\textwidth}{\includegraphics[width=0.315\textwidth]{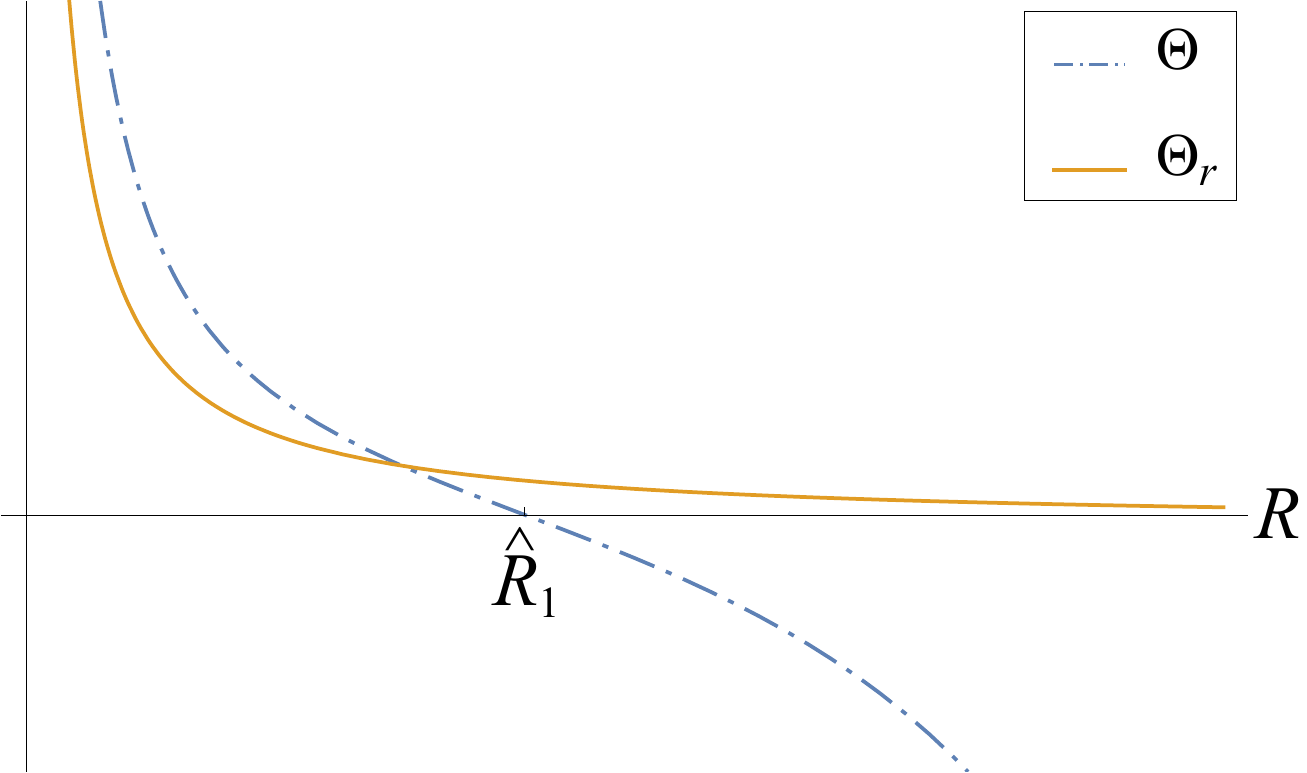}\\(a)} \ \
\parbox[c]{0.32\textwidth}{\includegraphics[width=0.315\textwidth]{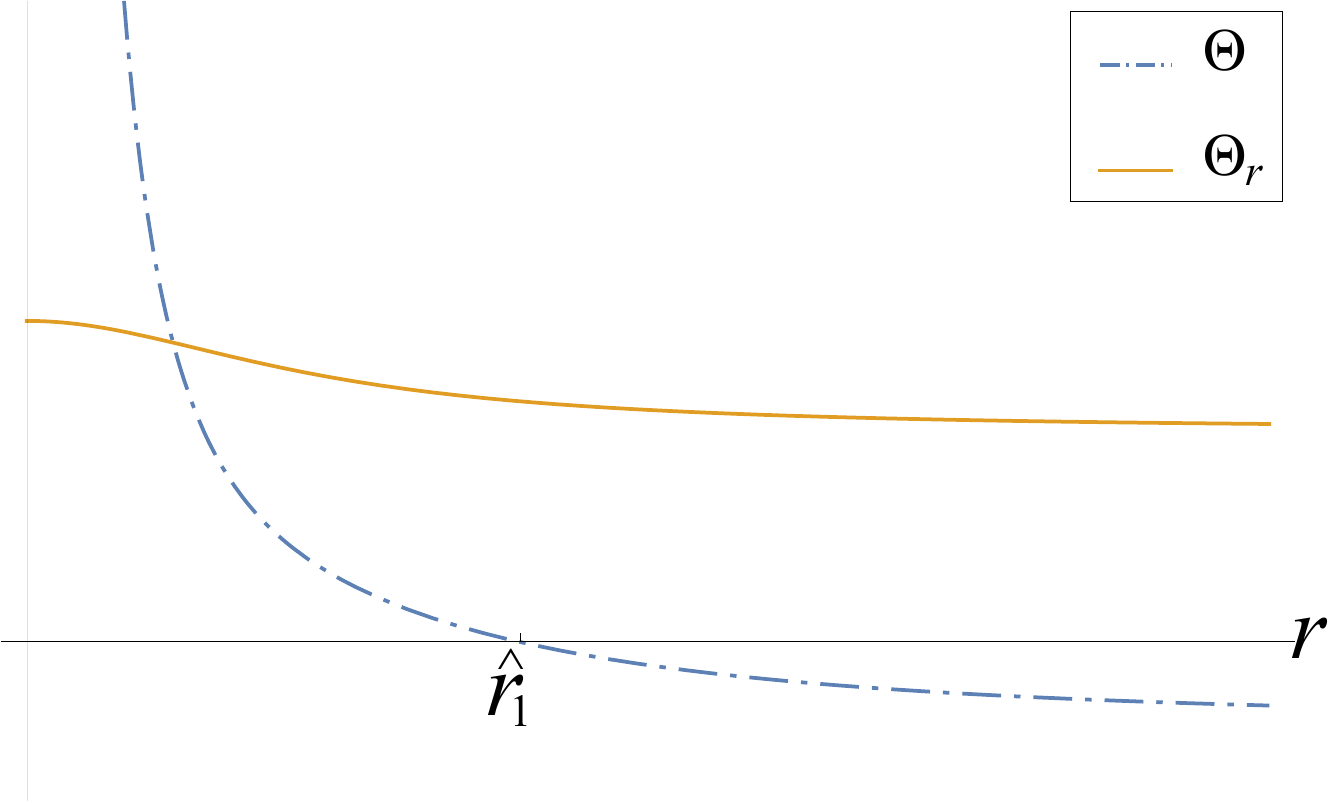}\\(b)} \ \
\parbox[c]{0.32\textwidth}{\includegraphics[width=0.315\textwidth]{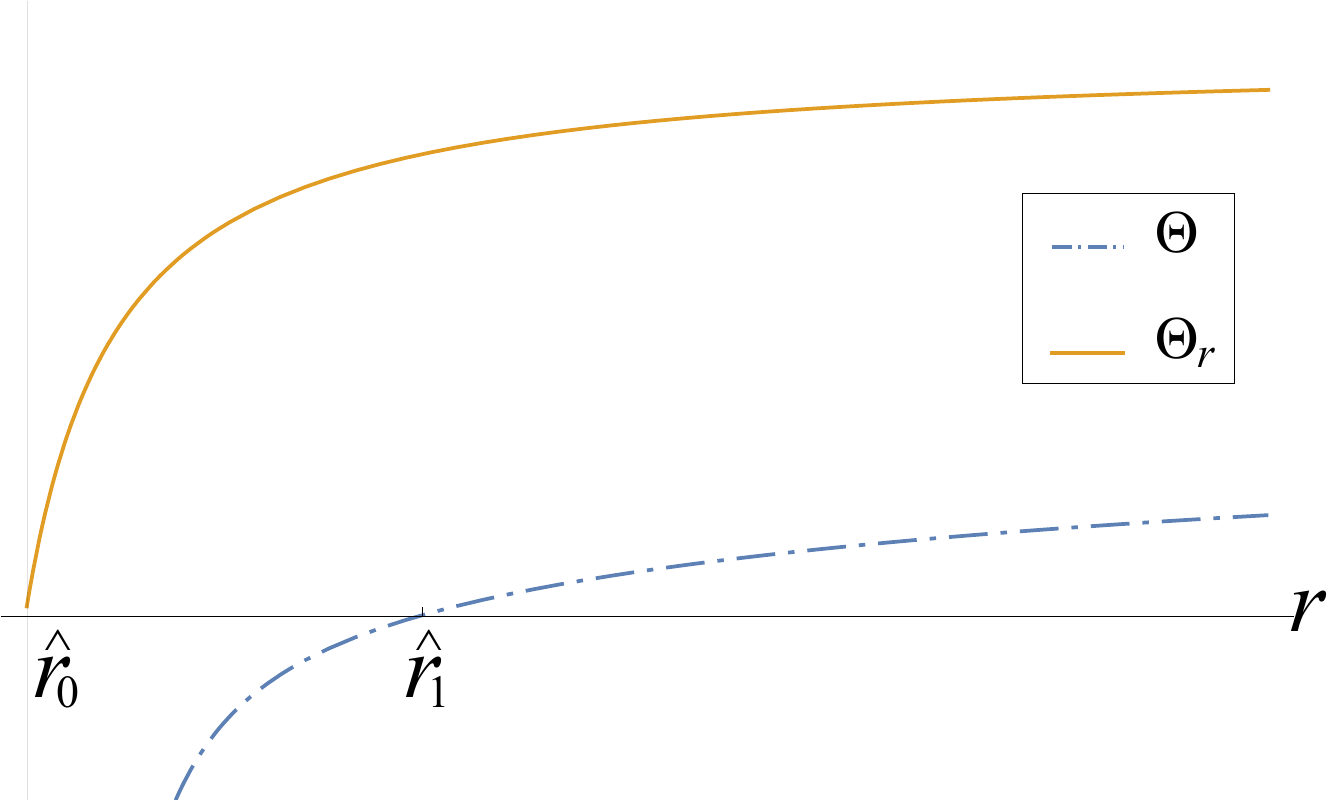}\\(c)}} 
\caption{(a) $R$-dependence of the temperatures $\Theta$, of the source ideal gas, and $\Theta_r$, of the test radiation fluid for a fixed $r_1$. The radial profile of these temperatures for a fixed $R_1$ are plotted in (b) (case I with $R_1 > - k_1/(4 k_2)$ and case II) and (c) (case III).} 
\label{Fig-6}
\end{figure*}
%

%%%%%%%%%%%%%%%%%%%%%%%%%%%%%
\subsection{SM: $R$-dependence and radial profiles}
\label{subsec-Model-Nos-Evolution}
%%%%%%%%%%%%%%%%%%%%%%%%%%%%%%%%%%%%%%%%%%%%%%%%%%%%%%

From now on we only consider the regions where the singular models approach a Synge gas (dark blue regions in Fig. \ref{Fig-2}). We have then three cases: I) ${\cal R}_0^{<}$ ($k_2 < 0$) with $k_1 \geq 0$, II) ${\cal R}_0^{<}$ ($k_2 < 0$) with $k_1 < 0$, and III) ${\cal R}_1^{>}$ ($k_2 > 0$ and $k_1 < 0$). 

In any case, for a suitable fixed value of the radial coordinate $r_1$, the hyperbola $\pi=0$ contains a point $(r_1^2, \hat{R}_1)$. In fact, the real function $\pi_1(R) \equiv \pi(r_1^2, R)$ decreases in the interval $[0, \hat{R}_1]$ between $1/3$ and zero (see Fig. \ref{Fig-3}(a)). Consequently the pressure $p_1(R) = p(r^2_1, R)$ is a decreasing function that vanishes at $\hat{R}_1$ (see Fig. \ref{Fig-3}(b)). 

In the Stephani universes the 3-space curvature depends on time, $\kappa= \kappa(R)$, and expression (\ref{curvature-ss}) implies that its sign depends on the sign of the function $k(R)= k_1 + k_2 R$. When $k_1 k_2 \leq 0$, the curvature vanishes at $R_{\kappa} = - k_1/k_2$. In case I, $k_2 < 0$, $k_1 \geq 0$, the curvature is a decreasing function and $R_{\kappa} < \hat{R}_1$ (region ${\cal R}_0^{<}$) for $r$ such that $r^2 < 4/(3 k_1)$. In case II, $k_2 < 0$, $k_1 <0$, the curvature $\kappa$ is always negative. And in case III, $k_2 > 0$, $k_1 <0$, the curvature is an increasing function and $R_{\kappa}> R_1$; consequently it is negative on the physical region ${\cal R}_1^{>}$ (see Fig. \ref{Fig-3}(c)). 

Given a fixed value $R_1$ of the function $R$, the radial profiles of the thermodynamic quantities also depend on the three different considered cases (see Fig. \ref{Fig-4}). In case I (region ${\cal R}^{<}_0$ with $k_1 \geq 0$), if $R_1 \leq - k_1/(4 k_2)$, the hydrodynamic functions $\pi$ and $\chi$ are decreasing functions which take values between $1/3$ and a non-negative real number (Fig. \ref{Fig-4}(a)). In case II (region ${\cal R}^{<}_0$ with $k_1 < 0$), or in case I with $R_1 > - k_1/(4 k_2)$, $\pi$ and $\chi$ are also decreasing functions which take the value $1/3$ at $r=0$ and vanish at a finite $r= \hat{r}_1$ and $r= \bar{r}_1$ (Fig. \ref{Fig-4}(b)).  Finally, in case III (region ${\cal R}_1^>$), $\pi$ and $\chi$ are increasing functions which are positive for $r>\hat{r}_1$ and $r > \bar{r}_1$ (Fig. \ref{Fig-4}(c)).

On the other hand, Fig. \ref{Fig-5} shows, also for a fixed $R_1$, the radial profile of the energy density $\rho$ (constant), pressure $p$ and matter density $n$. Again, the behavior is different for the three aforementioned cases. Fig. 5(a): for $r>0$, $p$ is decreasing and $n$ increasing, both positive. Fig. 5(b): for $r<\hat{r}_1$, $p$ and $n$ have the same behavior, and $p(\hat{r}_1)=0$ and $n=\rho$ at $r=\hat{r}_1$. Fig. 5(c): in this case $p$ is increasing and $n$ decreasing for $r> \hat{r}_1$.

Fig. \ref{Fig-6} describes the behavior of both the temperature $\Theta$ of the source ideal gas and the temperature $\Theta_r$ of the test radiation fluid. For a fixed $r_1$, both temperatures decrease with $R$, and $\Theta$ can vanish at $\hat{R}_1$ (see Fig. \ref{Fig-6}(a)). For a fixed $R_1$ the radial profile depends on the model. In Fig. \ref{Fig-6}(b) we have plotted the cases I with $R_1 > - k_1/(4 k_2)$ and II, where both temperatures decrease and $\Theta$ vanishes at $r=\hat{r}_1$. And Fig. \ref{Fig-6}(c) shows case III, where both temperatures increase and $\Theta$ is positive for $r>\hat{r}_1$.

\begin{figure*}
\parbox[c]{0.43\textwidth}{\includegraphics[width=0.43\textwidth]{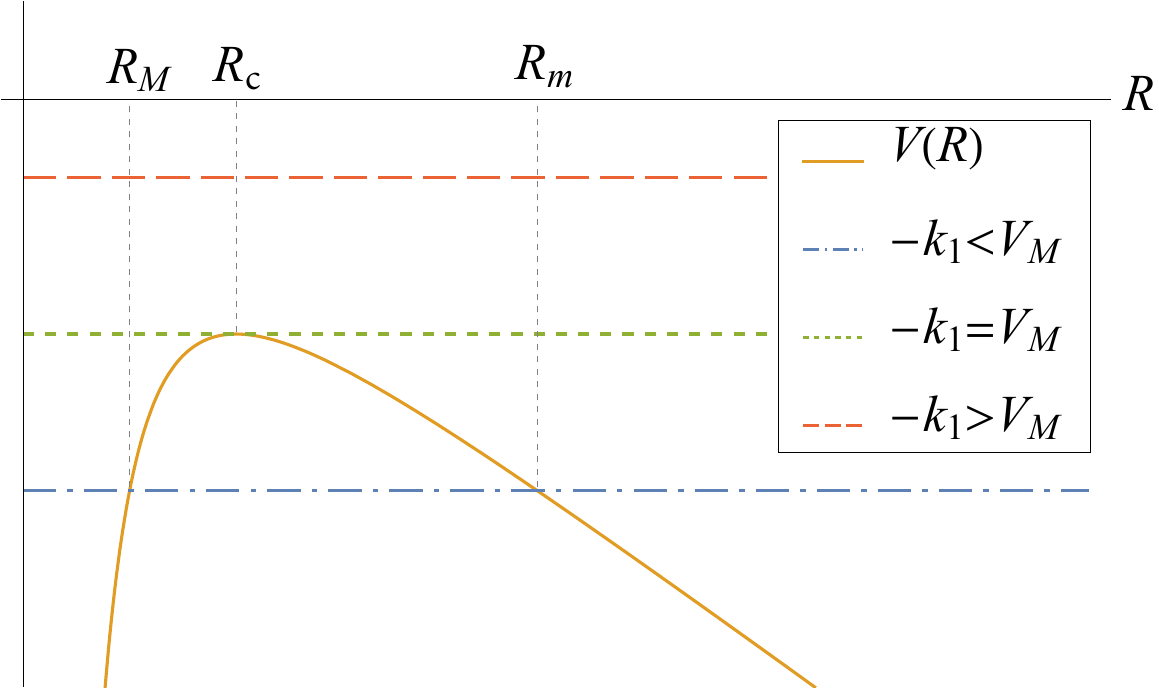}} \qquad
{\parbox[c]{0.43\textwidth}{\includegraphics[width=0.43\textwidth]{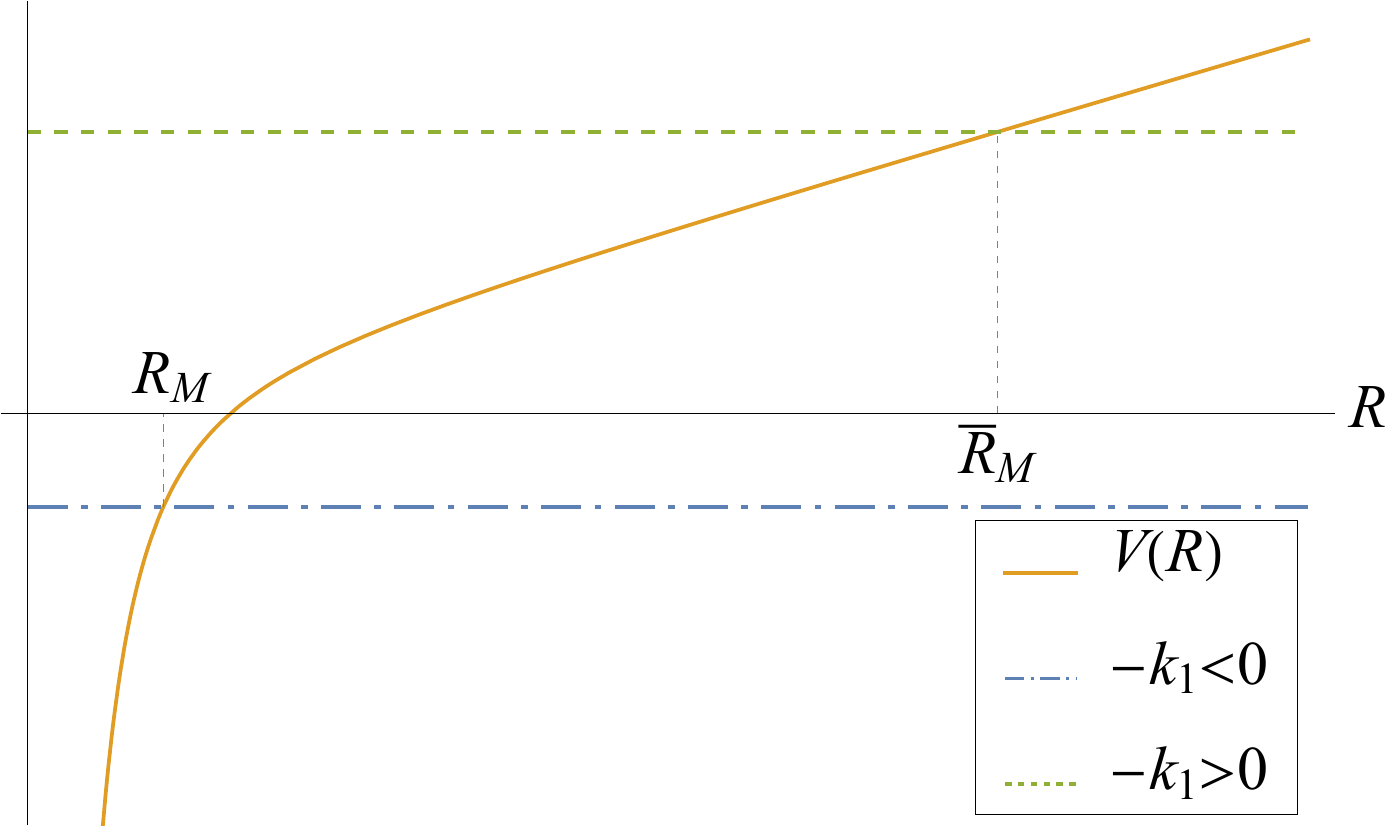}} \\[1mm]
\parbox[c]{0.43\textwidth}{\includegraphics[width=0.43\textwidth]{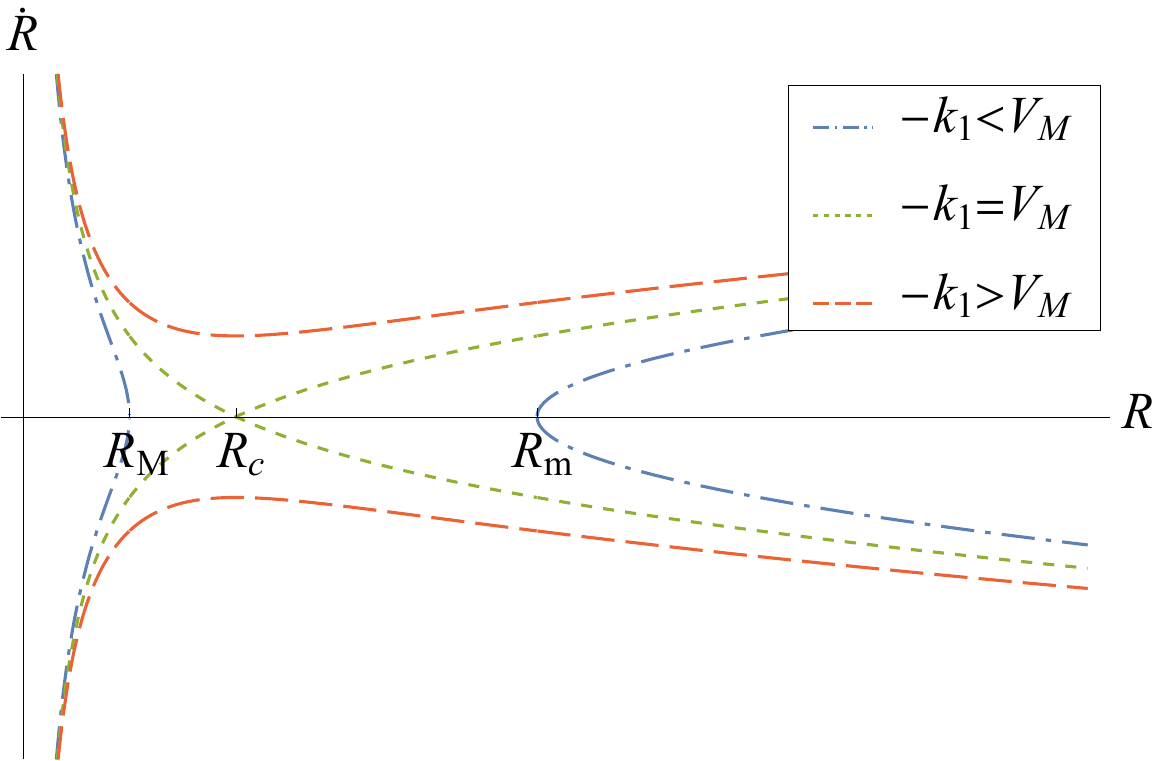}} \qquad
\parbox[c]{0.43\textwidth}{\includegraphics[width=0.43\textwidth]{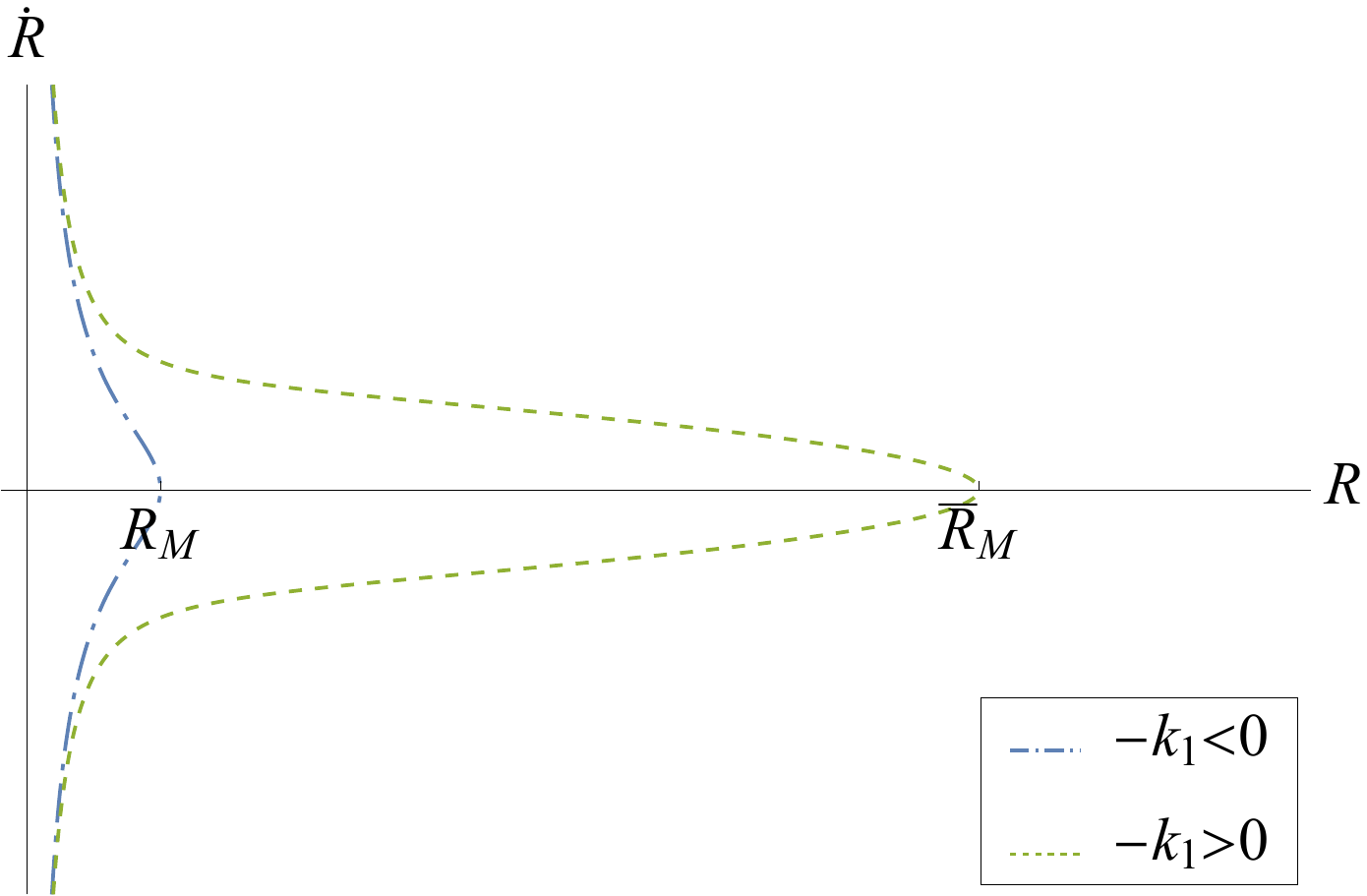}} \\[0mm]
\centerline{$k_2< 0$ \qquad \qquad \qquad  \qquad \qquad \qquad \qquad \qquad \qquad  \qquad $k_2 >0$}}
\caption{Trajectories in the phase space $\{R, \dot{R}\}$, which determine the behavior of the metric function $R(t)$. When $k_2 < 0$ (left panels), we could have closed, asymptotic and open models; when $k_2 >0$ (right panels), there are only closed models.} 
\label{Fig-7}
\end{figure*}
%

%%%%%%%%%%%%%%%%%%%%%%%%%%%%%
\subsection{SM: The generalized Friedmann equation}
\label{subsec-Model-Nos-Friedmann}
%%%%%%%%%%%%%%%%%%%%%%%%%%%%%%%%%%%%%%%%%%%%%%%%%%%%%%

For the singular models (\ref{singular-rho}), the generalized Friedmann equation (\ref{Friedmann-ideal-radiacio}) can be written as:
\begin{equation}
 \dot{R} = \sqrt{\!-k_1\! -\! V(R)}, \qquad V(R) \equiv k_2 R - \frac{\tilde{\rho}_0}{R^2} , 
\end{equation}
where $\tilde{\rho}_0 = \frac13 \rho_0 R_0^4$. Then, we can study the qualitative behavior of the function $R(t)$ by drawing the effective potential $V(R)$ and analyzing the trajectories in the phase plane $\{R, \dot{R}\}$ (see Fig. \ref{Fig-7}). 

Depending on the sign of $k_2$, we have two qualitatively different effective potentials. Now, we analyze the three cases I, II and III considered in subsection above. 

In case II (region ${\cal R}_0^{<}$ with $k_1 < 0$), we have $k_2<0$ (left panels in Fig. \ref{Fig-7}). Moreover, $-k_1 > V_M$, and the solution is valid for $r<r_{\infty}$. Then, we obtain an accelerated expanding model for values of $R$ larger than a critic value $R_c$, and for any $r_1<r_{\infty}$ the pressure vanishes at a finite time $t_1$ ($R(t_1))= R_1$).

In case I (region ${\cal R}_0^{<}$ with $k_1 \geq 0$), we have $k_2<0$ (left panels in Fig. \ref{Fig-7}), and three different models can occur. If $-k_1 > V_M$, we obtain a model similar to that of case II but now valid for any $r>0$. If $-k_1= V_M$, we obtain an asymptotic expanding model with $R \rightarrow R_c$; for small values of $r$, the pressure never becomes zero, but for large values of $r$, the pressure vanishes at $R_1 < R_c$. Finally, if $-k_1 < V_M$, we have closed models, with a maximum value of $R$, $R_M < R_{\kappa}$; generically, a $\bar{r}$ exists such that $R_M < \hat{R}_1$ if $r < \bar{r}$ and $R_M > \hat{R}_1$ if $r > \bar{r}$; $\bar{r} = \infty$ for large values of $k_1$. 

In case III (region ${\cal R}_1^{>}$ with $k_1 < 0$), we have $k_2>0$ (right panels in Fig. \ref{Fig-7}), and we also obtain closed models, but the pressure vanishes before the contracting era, $\hat{R}_1 < R_M$. Moreover, $\hat{R}_1 < R_{\kappa}$.

%%%%%%%%%%%%%%%%%%%%%%%%%%%%%
\subsection{SM: Physical interpretation and further prospects}
\label{subsec-Model-Nos-Friedmann}
%%%%%%%%%%%%%%%%%%%%%%%%%%%%%%%%%%%%%%%%%%%%%%%%%%%%%%

To sum up, the solutions considered in this section model a spherically symmetric spacetime inhomogeneity caused by an ultrarelativistic gas with homogeneous energy density, and inhomogeneous pressure, matter density and temperature. This inhomogeneity is compatible with a decoupled test inhomogeneous radiation fluid, which is isotropic as measured by the observer comoving with the matter fluid. 

Wide ranges of parameters lead to physically realistic models. All of them start with a hot ultrarelativistic fluid that cools down with time, in most cases even becoming dust. Their radial profiles, however, depend on the considered model. Most of them are only physically admissible up to or from a certain value of the radial coordinate, where the pressure vanishes. In some cases we have a void of hot matter (cases I and II), and in other cases the matter density decreases, and the temperature increases, with $r$ (case III).

Thus, the models are useful to describe local inhomogeneities. Nevertheless, in order to be useful globally, they must be matched with other dust models beyond the hypersurface $p=0$.

%%%%%%%%%%%%%%%%%%%%%%%%%
\section{Discussion}
\label{sec-conclusions}
%%%%%%%%%%%%%%%%%%%%%%%%%

In this paper we have studied general properties of the thermodynamic Stephani universes, and we have analyzed the constraints that some specific physical requirements impose on the models.

We have focused on the solutions where the observer comoving with the fluid flow can measure a state of isotropic radiation. The models that we consider highlight a long-known fact (see \citep{FMP-isotropa, Clarkson-1999} and references therein): an inhomogeneous perfect fluid solution can be compatible with an observed inhomogeneous and isotropic radiation.

Although our purpose here has not been to look for cosmological models compatible with the observational data, our study shows that some of our models, or other similar ones that could be obtained with an analogous approach, could model local nonlinear inhomogeneities of the real Universe.

The results that we have obtained suggest many open problems whose study goes beyond the scope of this work. Regarding the specific models studied here, we can quote the following further work: (i) for the singular models considered in Sec. \ref{sec-Model-Nos}, to match our solutions with a dust model through the junction surface $\pi(R,r)=0$; (ii) to make an accurate analysis of the parameters of the models to achieve the more suitable values for physically realistic models; (iii) to investigate the regular model in detail as we have done with the singular one.

In the inhomogeneities observed in the real Universe, matter moves with respect to the cosmological observer who observes an almost isotropic background radiation. To study the radial profiles and the evolution of such nonlinear inhomogeneities we are interested in obtaining solutions with test isotropic radiation for a cosmological observer and a perfect fluid source with a noncomoving flow.  

A further study to be made consists in analyzing in depth the flow of the thermodynamic Stephani universes taking into account the kinematic approaches presented in \cite{FS-KCIG} and \cite{SMF-K}. This study will enable us to determine other test fluids which are comoving with the Stephani cosmological observer.

%%%%%%%%%%%%%%%%%%%%%%%5

\begin{acknowledgements}
This work has been supported by the Generalitat Valenciana Project CIAICO/2022/252, the Spanish Ministerio de Ciencia, Innovaci\'on, Project PID2019-109753GB-C21/AEI/10.13039/501100011033, and the Plan Recuperaci\'on, Transformaci\'on y Resiliencia, project ASFAE/2022/001, with funding from European Union NextGenerationEU (PRTR-C17.I1). S.M. acknowledges financial support from the Generalitat Valenciana (grant CIACIF/2021/028). 
\end{acknowledgements}

%%%%%%%%%%%%%%%%%%%%%%%%%%%%%%%%%%%%%%%%%%%%%%%%%%%%%%%%%%%%%%%

\appendix

\section{} \label{AppendixA}
The metric of the thermodynamic Stephani universes with spherical symmetry is given by $(10-12)$ with $\varepsilon = 1$. If we perform the following change of spatial coordinates:
	\begin{eqnarray}
		x = \frac{x'}{H}  , \quad y = \frac{y'}{H}  , \quad z = \frac{z' -1 + \frac12 r'^2}{H}  , \  \\
		r'^2= x'^2 + y'^2 + z'^2  , \quad H = 1 + z' + \frac14 r'^2 ; \
	\end{eqnarray}
and we define
	\begin{equation} \label{funcions barra}
		\bar{R} \equiv \frac{2 R}{1 - 2 b} \, , \qquad k \equiv 4 \frac{1 + 2b}{1 - 2b} \, ;
	\end{equation}
and perform the change of temporal coordinate from $t$ to $t'$ such that
	\begin{equation}
		\theta (t') = 3 \frac{\dot{\bar{R}}}{\bar{R}} \, ;
	\end{equation}
the metric of the thermodynamic Stephani universes with spherical symmetry can be written as
	\begin{eqnarray} 
		\dif s^2 = -\bar{\alpha}^2 \dif t'^2 + \bar{\Omega}^2 (\dif x'^2 + \dif y'^2 + \dif z'^2), \\
	\label{spherical alpha and Omega}
		\bar{\Omega} \equiv \frac{\bar{R}(t)}{1 + \frac14 k \, r'^2} \, , \qquad  \qquad \bar{\alpha} \equiv \bar{R} \, \partial_{\bar{R}} \ln\bar{\Omega} \, . 
	\end{eqnarray}
With that, the hydrodynamic quantities $\rho$ and $p$ and the curvature of the spatial synchronizations take the following expressions:
	\begin{equation}
		\rho = \frac{3}{\bar{R}^2}(\dot{\bar{R}}^2 +k)  ,  \quad p = -\rho - \frac{\bar{R}}{3}\frac{\partial_{\bar{R}} \rho}{\bar{\alpha}}, \quad \kappa(t) = \frac{k}{\bar{R}^2}  . \
	\end{equation}
%

%%%%%%%%%%%%%%%%%%%%%%%%%%%%%%%%%%%%%%%%%%%%%%%%%%%%%%%%%

\bibliography{PRD}

\end{document}